\documentclass[runningheads]{llncs}
\usepackage{url}
\usepackage{xspace}
\usepackage{microtype} 
\usepackage{todonotes}
\usepackage[inline]{enumitem}
\usepackage{subcaption}\usepackage{amsmath}

\usepackage[unicode,colorlinks=true,allcolors=teal]{hyperref}

\usepackage{thmtools}
\usepackage{thm-restate}

\newcommand{\RS}{RS\xspace}
\newcommand{\CEP}{CE\xspace}
\newcommand{\CR}{CR\xspace}
\newcommand{\CO}{CO\xspace}
\newcommand{\ERS}{ERS\xspace}

\renewcommand{\orcidID}[1]{\href{https://orcid.org/#1}{\includegraphics[scale=.03]{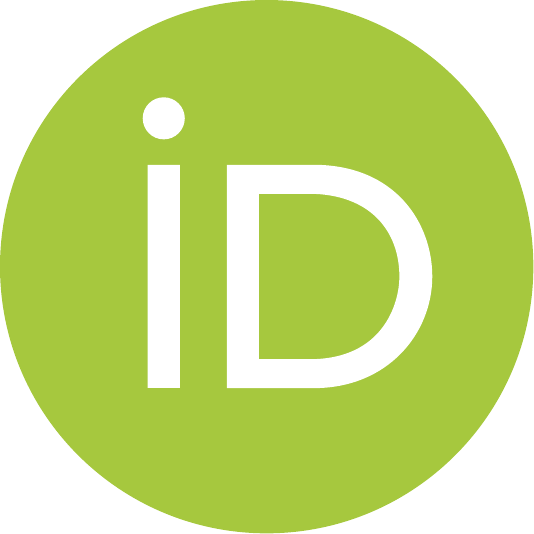}}}

\title{Different Types of Isomorphisms of Drawings of Complete Multipartite Graphs\thanks{O.A.\, and B.V.\, partially supported by Austrian Science Fund (FWF) within the collaborative DACH project \emph{Arrangements and Drawings} as FWF project \mbox{I 3340-N35}. O.A.\, B.V.\,, and A.W.\, partially supported by FWF grant~W1230. 
\newline We thank the reviewers of EuroCG'21 and GD'23 for their very helpful comments.}}
\titlerunning{Types of Isomorphisms of Drawings of Complete Multipartite Graphs}
\author{Oswin Aichholzer\orcidID{0000-0002-2364-0583} 
	\and Birgit Vogtenhuber\orcidID{0000-0002-7166-4467} 
	\and Alexandra~Weinberger\orcidID{0000-0001-8553-6661}} 
\institute{Institute of Software Technology, Graz University of Technology, Austria
	\texttt{oaich|bvogt|weinberger@ist.tugraz.at}}
\authorrunning{O. Aichholzer, B. Vogtenhuber, A. Weinberger}

\begin{document}

\maketitle
\begin{abstract}
		Simple drawings are drawings of graphs in which any two edges intersect at most once (either at a common endpoint or a proper crossing),
		and no edge intersects itself.
		We analyze several characteristics of simple drawings of complete multipartite graphs: which pairs of edges cross, in which order they cross, and the cyclic order around vertices and crossings, respectively.
		We consider all possible combinations of how two drawings can share some characteristics and determine which other characteristics they imply and which they do not imply.
		Our main results are that for simple drawings of complete multipartite graphs, the orders in which edges cross determine all other considered characteristics.
		Further, if all partition classes have at least three vertices, then the pairs of edges that cross determine the rotation system and the rotation around the crossings determine the extended rotation system. 
		We also show that most other implications -- including the ones that hold for complete graphs -- do not hold for complete multipartite graphs. 
		Using this analysis, we establish which types of isomorphisms are meaningful for simple drawings of complete multipartite graphs.
	
\end{abstract}

\section{Introduction}
A \emph{simple drawing} of a graph is a drawing in the plane or on the sphere in which vertices are represented as points, edges are non-self-intersecting curves
connecting their endpoints and not passing through any other point representing a vertex. Further, every pair of edges intersects at most once (either in a common endpoint or in a proper crossing). 
The \emph{rotation} of a vertex or crossing in a labeled drawing is the clockwise cyclic order of the endpoints of incident edges around this vertex or crossing. (We remark that for crossings, we always note as endpoints the vertices induced by the respective edge-fragments.) 
To describe simple drawings, the following are the most commonly used properties.

\begin{itemize}
	\item 
	The collection of the rotations of all vertices. This collection 
		is called the \emph{rotation system} [RS];
	\item
	The pairs of edges that cross [CE];
	\item
	The collection of the rotations of all crossings [CR];
	\item
	The collection of the rotations of all vertices and all crossings. This collection is called the \emph{extended rotation system} [ERS];
	\item
	The collection of the \emph{crossing orders} of all edges, that is, along each edge, the order in which the edge crosses other
	edges [CO];
\end{itemize} 

For each such property, we can define a type of isomorphism. Two labeled simple drawings of the same graph are \ldots
\begin{itemize}[label*=\ldots]
	\item
	\emph{RS-isomorphic} if either for each vertex the rotation is the same in both drawings of for each vertex the rotation is inverse between the two drawings.
	\item 
	\emph{CE-isomorphic} if the same pairs of edges cross. In other literature (e.g.~\cite{kyncl_extended_rot}), this property is also called \emph{weak isomorphism}.
	\item
	\emph{CR-isomorphic} if either for each crossing the rotation is the same in both drawings or for each crossing the rotation is inverse between the two drawings.
	\item
	\emph{ERS-isomorphic} if their extended rotation systems are the same or inverse (where inverse 
		means for each vertex and each crossing the rotation in one drawing is the inverse from the rotation in the other).
\item
\emph{CO-isomorphic} if for each edge, its crossing order is the same in both drawings.
	\item 
	\emph{strongly isomorphic} if there exists a homeomorphism of the sphere such that one drawing is mapped to the other.
\end{itemize} 

Unlabeled simple drawings are isomorphic with respect to some type of isomorphism if there exists a labeling such that the labeled drawings are isomorphic with respect to that type. In this paper, when we say that two simple drawings are isomorphic without specifying, the statement holds for both the labeled and the unlabeled case (where often the labeled case follows from the unlabeled case). 

\begin{figure}[htb] 
	\centering
	\begin{subfigure}{0.45\textwidth}
	\centering
	\includegraphics[page=5]{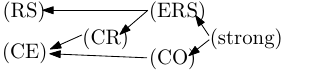}
	\caption {Implications for any graph. }
	\label{fig:def_imp}
	\end{subfigure}\hfill 
	\begin{subfigure}{0.45\textwidth}
	\centering
	\includegraphics[page=6]{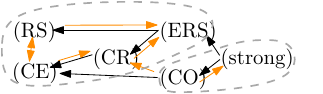}
	\caption {Implications for~$K_n$ with $n\geq 6$.}
	\label{fig:kn_imp}
\end{subfigure}
	\caption {Implications between isomorphisms. Black arrows hold by definition. Orange arrows hold for $K_n$ with $n\geq 6$. Dashed curves group the isomorphism classes by equivalences (for $K_n, n\ge 6$).}
\end{figure}

Some of the isomorphisms imply other isomorphisms for any graph by definition; see Figure~\ref{fig:def_imp} for a depiction of the trivial implications that hold for all graphs. The extended rotation system combines the information of the rotation system with the information of the rotations around the crossings. Thus, if two drawings are ERS-isomorphic, they are also CR-isomorphic and RS-isomorphic. Since the crossings have to be the same for the rotations around the crossings to be the same (and also for the order of crossings to be the same), each of ERS-isomorphism, CR-isomorphism, and CO-isomorphism implies CE-isomorphism.

Further, Kyn\v{c}l~\cite{kyncl_improved} 
showed the following useful 
characterization of strong isomorphism (which we restate here with the above-defined terminology).
\begin{theorem}\cite{kyncl_improved}\label{thm:kyncl}
	Two connected, labeled drawings~$D$ and~$D'$ of the same graph on the sphere are \emph{strongly isomorphic} if and only if the following properties	hold simultaneously: 
\begin{enumerate*}[,label={(\roman*)}]
			\item\label{prop:order} The drawings~$D$ and~$D'$ are CO-isomorphic.  
			\item\label{prop:rotation} The drawings~$D$ and~$D'$ are ERS-isomorphic.
\end{enumerate*}
\end{theorem}

\subsubsection*{Complete Graphs.}
For drawings of complete graphs, more implications and also equivalences are known; see Figure~\ref{fig:kn_imp}.  
Concretely, RS-isomorphism, CE-isomorphism, and ERS-isomorphism (and thus also CR-isomorphism) are all equivalent to each other~\cite{gioan,gioan_final,kyncl_extended_rot}.
Further, as CO-isomorphism implies CE-isomorphism (and thus 
ERS-isomorphism), by Theorem~\ref{thm:kyncl}, CO-isomorphism implies strong isomorphism. 
For $n\leq 5$, 
the order of crossings along the edges can be derived from the pairs of crossing edges~\cite{gioan_proof,gioan,gioan_final,schaefer21}.
For $n \geq 6$ this is no longer the case; but 
any two simple drawings that are CE-isomorphic can be transformed into each other by 
a sequence of local operations called triangle flips (a.k.a.~Reidemeister moves of type III)~\cite{gioan_proof,gioan,gioan_final,schaefer21}. In conclusion, there are only two relevant classes of isomorphisms: 
One contains RS-, CE-, CR-, and ERS-isomorphism, as well as any combination of them;  the other contains strong and CO-isomorphism (plus any combination of the others).
We remark that $K_n$ coincides with the complete multipartite graph on $n$ vertices that has $n$ partition classes, each of them containing exactly one vertex.

\subsubsection*{Complete Multipartite Graphs.}
The main goal of this paper is to classify which types of isomorphism are relevant for drawings of complete multipartite graphs. 
Many of the implications that hold for the complete graph do not hold for non-complete graphs, including complete multipartite graphs.
For example, it is known that 
for simple drawings of complete multipartite graphs, \mbox{RS-isomorphism} does not imply CE-isomorphism; see Figure~\ref{fig:onlyROT_big}.
On the positive side, it still holds that two simple drawings of a complete multipartite graph that are ERS-isomorphic can be transformed into each other 
by a sequence of triangle flips~\cite{gioan_multi}. 
However, to the best of our knowledge, except this latter property and the implications that follow directly from the definitions, 
no implications between the different properties of simple drawings of (general) complete multipartite graphs 
have been known. 

\begin{figure}[htb]
	\centering
	\begin{subfigure}{0.45\textwidth}
	\centering
	\includegraphics[page=10]{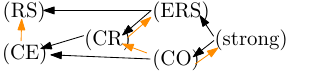}
	\caption {Complete multipartite graphs with at least three vertices in each class.}
	\label{fig:kmn_imp_a}
	\end{subfigure}\hfill 
	\begin{subfigure}{0.45\textwidth}
	\centering
	\includegraphics[page=11]{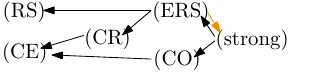}
	\caption {$K_{2,n}$ with $n \geq 2$. \linebreak \hfill}
	\label{fig:kmn_imp_b}
	\end{subfigure}
	\caption {Implications that hold for simple drawings of complete multipartite graphs. The implications which don't follow by definition are drawn orange.}
	\label{fig:kmn_imp}
\end{figure}

In this paper, we give a complete characterization which implications do or do not hold for drawings of complete multipartite graphs, depending on the cardinalities of their partition classes; cf.~Figure~\ref{fig:kmn_imp}. 
To obtain this characterization, we mostly focus on drawings of complete multipartite graphs in which every partition class consists of at least three vertices, and drawings of $K_{1,n}$ and $K_{2,n}$.   

\subsubsection*{Further Related Work.}
The majority of previous work on simple drawings is focused on the special case of complete graphs. Most of the work on more general complete multipartite graphs is focused on how to draw these graphs with as few crossings as possible. In 1954, Zarankiewicz~\cite{zaran} gave a (straight-line) drawing construction for complete bipartite graphs that is still conjectured to reach the minimum number of crossings over all simple drawings of such graphs. 
In 1971, Harborth~\cite{harborth_crossing} extended this result and gave a simple drawing of complete multipartite graphs that is conjectured to reach the minimum number of crossings (and can be drawn straight-line if there are at most three partition classes). 
For further work on the crossing number of complete multipartite graphs see also \cite{special_crossing_number2,EGCcrossingnumber,gethner2017crossing,special_crossing_number3,special_crossing_number1,crossing_number_survey} and references therein.

Beyond the crossing number problem, Cardinal and Felsner~\cite{CardinalFelsner2018} studied rotation systems of complete bipartite graphs and their realization as special simple drawings. Further, there has been work on plane subdrawings~\cite{gd_shootingstars,mengersen_phd}, and on triangle flips~\cite{gioan_multi} in simple drawings of complete multipartite graphs, as well as on the enumeration of simple drawings of $K_{m,n}$ with $m\leq 3,n\leq3$~\cite{harborth_enumeration}.

\subsection{Obtained Results} 
\begin{figure}
	\centering
	\begin{subfigure}{0.45\textwidth}
	\centering
	\includegraphics[page=14]{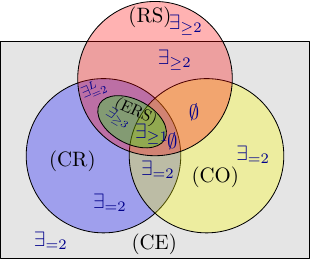}
	\caption {Complete multipartite graphs.}
	\label{fig:genKn_a}
	\end{subfigure}\hfill 
	\begin{subfigure}{0.45\textwidth}
	\centering
	\includegraphics[page=15]{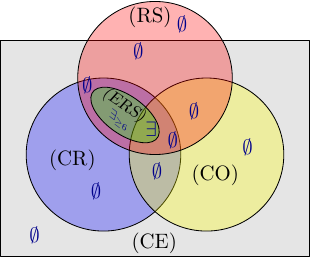}
	\caption {Complete graphs.}
	\label{fig:genKn_b}
	\end{subfigure}
	\caption {Classification of all possible combinations of different properties. The used notation is explained in the text above.}
\label{fig:genKn}
\end{figure}

An overview of our results is given in Figure~\ref{fig:genKn_a}, where we use the following symbols:  
In areas marked with~$\emptyset$, there are no two labeled or unlabeled simple drawings of complete multipartite graphs such that the two drawings share exactly the intersecting properties (and do not share any other properties).
In the areas marked with~$\exists_{=2}^L$, 
no labeled or unlabeled simple drawings of complete multipartite graphs with at least three vertices in each partition class 
such that the drawings share exactly the intersecting properties, 
but there are labeled simple drawings of~$K_{2,n}$ that share exactly the intersecting properties; 
and in areas marked with~$\exists_{=2}$, there are labeled and unlabeled simple drawings of~$K_{2,n}$ that share exactly the intersecting properties.  
In areas marked with with~$\exists_{\geq x}$ for $x\in \{1,2,3\}$, there are labeled and unlabeled simple drawings sharing exactly the intersecting properties of complete multipartite graphs in which all partition classes have at least $x$  
vertices, but no such drawings if the smallest partition class has less than $x$ vertices. 
Especially, for $x=1$, there are such simple drawings independent of the sizes of the partition class sharing exactly the intersecting properties.
For comparison, Figure~\ref{fig:genKn_b} shows the analogous diagram for $K_{n}$. 
In this figure, the label $\exists_{\geq 6}$ means that there exist labeled and unlabeled drawings of $K_n$ with $n \geq 6$, that share exactly the intersecting properties, while $\exists$ means that there exist such drawings of $K_n$ for any~$n$.

We list our results by first stating implications between (combinations of) 
isomorphisms that hold, and ending with a list of
isomorphisms that do not imply each other.
We consider all combinations
except for implications that follow from our statements by definition.
All statements that do not explicitly specify whether the considered drawings are labeled or unlabeled hold for both settings.
Together, this gives a complete characterization which implications hold. 

\begin{restatable}{theorem}{orderRS}\label{thm:orderRS}
	Let $G$ be a complete multipartite graph. Then any two unlabeled simple drawings of $G$ that are
	RS-isomorphic and CO-isomorphic are strongly isomorphic. If $G$ has 
	at least five vertices, then also any two labeled simple drawings of $G$ that are RS-isomorphic and CO-isomorphic are strongly isomorphic.
\end{restatable}

\begin{restatable}{theorem}{thmrot}\label{thm:rot}
	Let $G$ be a complete multipartite graph in which each partition class has at least three vertices. Then any two simple drawings of $G$ 
	that are CE-isomorphic are also RS-isomorphic.
\end{restatable}

\begin{restatable}{theorem}{corcr}\label{thm:cr_ers}
Let $G$ be a complete multipartite graph in which each partition class has at least three vertices. Then any two simple drawings of $G$ that are CR-isomorphic are ERS-isomorphic.
\end{restatable}

\begin{restatable}{cor}{thmsemi}\label{thm:semi}
		Let $G$ be a complete multipartite graph in which each partition class has at least three vertices. Then any two simple drawings of $G$  that are CO-isomorphic are strongly isomorphic.
\end{restatable}

Simple drawings of~$K_{2,n}$ behave differently in many ways. 
While Theorem~\ref{thm:orderRS} still holds, the other three statements do not. 
This different behavior of simple drawings of~$K_{2,n}$ is in most parts due to the fact that for all $n$ vertices of the larger partition class, the rotation of the vertex does not contain any information because it is a cyclic order of two elements. 
On the other hand, 
ERS-isomorphism does now imply strong isomorphism. 
This follows from the before-mentioned result on triangle flips~\cite{gioan_multi} but can also be shown directly; see Appendix~\ref{sec:proof_K2n}. Figure~\ref{fig:separate} compares the implications 
for complete multipartite graphs in which all partition classes contain at least three vertices with the ones for~$K_{2,n}$. 
The combination marked with $\exists^L$ in Figure~\ref{fig:separate_b} exists at least for labeled drawings of~$K_{2,n}$. 

\begin{figure}[tb]
	\centering
	\begin{subfigure}{0.45\textwidth}
	\centering
	\includegraphics[page=16]{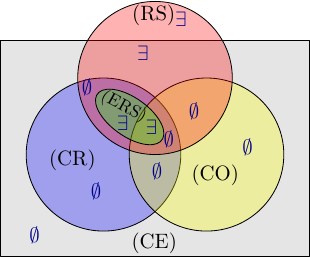}
		\caption {Complete multipartite graphs with at least three vertices in each class.}
	\label{fig:separate_a}
	\end{subfigure}\hfill 
	\begin{subfigure}{0.45\textwidth}
	\centering
	\includegraphics[page=17]{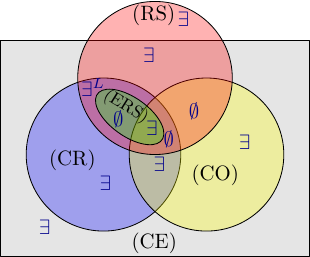}
	\caption {$K_{2,n}$.\linebreak \hfill}
	\label{fig:separate_b}
	\end{subfigure}
	\caption {Classification of all possible combinations of different properties. }
\label{fig:separate}
\end{figure}

Simple drawings of~$K_{1,n}$ again behave differently since all such drawings are plane. Thus, for~$K_{1,n}$, RS-isomorphism implies strong isomorphism for labeled drawings, and all unlabeled simple drawings 
are strongly isomorphic.       

We now turn to types of isomorphisms and combinations of types that do not imply others. 

\begin{figure} 
	\centering
		\includegraphics[page=10]{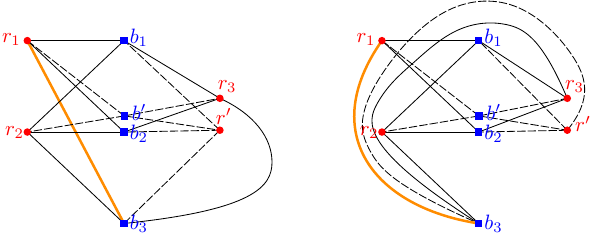}
	\caption {Simple drawings of~$K_{\textcolor{red}{3},\textcolor{blue}{3}}$ that are RS-isomorphic, but not CE-isomorphic. 
	The drawings can be extended via the vertices $\textcolor{blue}{b'}$, $\textcolor{red}{r'}$ and copies of them to~$K_{\textcolor{red}{m},\textcolor{blue}{n}}$ for $\textcolor{red}{m}\geq4,\textcolor{blue}{n}\geq4$. 
	}
\label{fig:onlyROT_big}
\end{figure}

\begin{figure} 
	\centering
		\includegraphics[page=7]{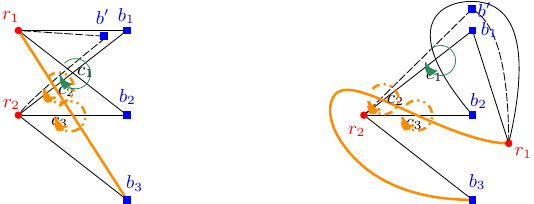}
		\caption {Two labeled simple drawings of~$K_{\textcolor{red}{2},\textcolor{blue}{3}}$ that are CE-isomorphic and RS-isomorphic, 
			but the crossings along the (bold, orange) edge $\textcolor{red}{r_1}$$\textcolor{blue}{b_3}$ are in different order and the rotations of crossings are different. 
	Crossings with a solid, green arc around them have the same crossing rotation in both drawings, while crossings with a dash-dotted, orange arc around them have inverse rotations between the drawings.}
	\label{fig:CEP_RS_K2n_small}
\end{figure}

\begin{figure} 
	\centering
		\includegraphics[page=11]{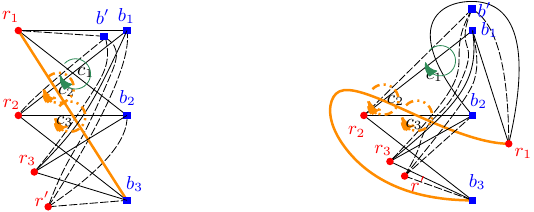}
	\caption {Two labeled simple drawings  of~$K_{\textcolor{red}{3},\textcolor{blue}{3}}$ that are CE-isomorphic and RS-isomorphic, 
		but the crossings along the (bold, orange) edge $\textcolor{red}{r_1}$$\textcolor{blue}{b_3}$ an edge ($\textcolor{red}{r_1}\textcolor{blue}{b_3}$ drawn in bold, orange) are in different order and the rotation around crossings is different.
}
	\label{fig:CEP_RS_K2n_big}
\end{figure}

\begin{figure} 
	\centering
	\includegraphics[page=9]{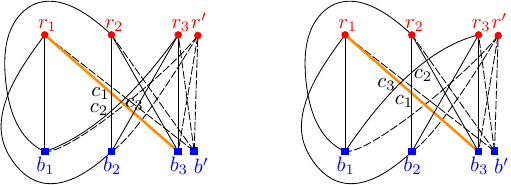}
	\caption {Two simple drawings of~$K_{\textcolor{red}{3},\textcolor{blue}{3}}$, which are ERS-isomorphic, but the crossings along the (bold, orange) edge $\textcolor{red}{r_1}$$\textcolor{blue}{r_3}$ is different.} 
	\label{fig:ERS_Kmn}
\end{figure}

\begin{figure} 
	\centering
	\includegraphics[page=1]{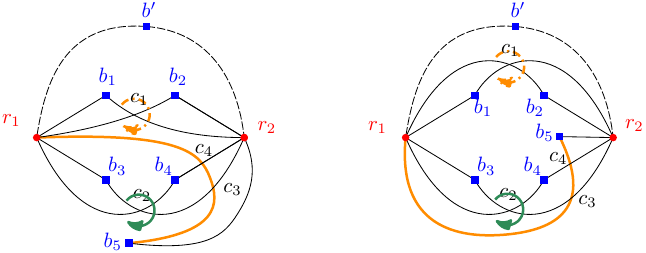}
	\caption {Two simple drawings of~$K_{\textcolor{red}{2},\textcolor{blue}{5}}$, which are CE-isomorphic, but neither RS-isomorphic nor CO-isomorphic nor CR-isomorphic.} 
	\label{fig:only_CEP_K2n}
\end{figure}

\begin{figure} 
	\centering
	\includegraphics[page=5]{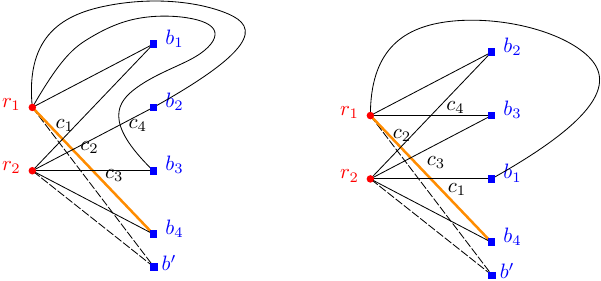}
	\caption {Two simple drawings of~$K_{\textcolor{red}{2},\textcolor{blue}{4}}$, which are CR-isomorphic, but neither RS-isomorphic nor CO-isomorphic.} 
	\label{fig:CEP_CR_K2n}
\end{figure}

\begin{figure} 
	\centering
	\includegraphics[page=4]{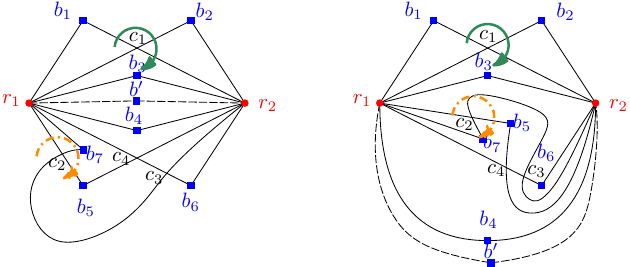}
	\caption {Two simple drawings of~$K_{\textcolor{red}{2},\textcolor{blue}{7}}$, which are CE-isomorphic and CO-isomorphic, but neither RS-isomorphic nor CR-isomorphic. 
	}
	\label{fig:CEP_CO_K2n}
\end{figure}

\begin{figure} 
	\centering
	\includegraphics[page=6]{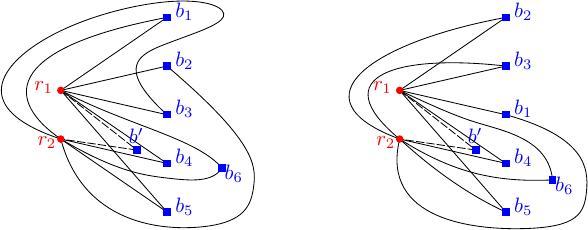}
	\caption {Two simple drawings of~$K_{\textcolor{red}{2},\textcolor{blue}{6}}$, which are CO-isomorphic and CR-isomorphic, but not RS-isomorphic. 
	}
	\label{fig:CEP_CR_CO_K2n}
\end{figure}

\begin{enumerate}[leftmargin=*,label={(\arabic*)}]
	\item There are simple drawings of $K_{m,n}$ that are RS-isomorphic but not CE-isomorphic; see Figures~\ref{fig:onlyROT} and~\ref{fig:onlyROT_big}.
	\item For $K_{m,n}$ with $m \geq 2$ and $n \geq 3$, there are simple drawings which are CE-isomorphic and RS-isomorphic but neither CR-isomorphic nor CO-isomorphic; see Figures~\ref{fig:CEP_RS_K2n_small} and~\ref{fig:CEP_RS_K2n_big}.
	\item For $K_{m,n}$ with $m \geq 3$ and $n \geq 3$, there are simple drawings which are ERS-isomorphic but not CO-isomorphic; see Figure~\ref{fig:ERS_Kmn}.
	\item For $K_{2,n}$, there are simple drawings which are CE-isomorphic but neither CO-isomorphic nor CR-isomorphic nor RS-isomorphic; see Figure~\ref{fig:only_CEP_K2n}.
	\item For $K_{2,n}$, there are simple drawings which are CR-isomorphic but neither CO-isomorphic nor RS-isomorphic; see Figure~\ref{fig:CEP_CR_K2n}. 
	\item For $K_{2,n}$, there are labeled simple drawings which are CR-isomorphic and RS-isomorphic but not ERS-isomorphic or CO-isomorphic; see Figure~\ref{fig:CR_RS_K2n}. 
	\item For $K_{2,n}$, there are simple drawings which are CO-isomorphic but not CR-isomorphic or RS-isomorphic; see Figure~\ref{fig:CEP_CO_K2n}. 
	\item For $K_{2,n}$, there are simple drawings which are CE-isomorphic, CR-isomorphic, and CO-isomorphic but not RS-isomorphic; see Figure~\ref{fig:CEP_CR_CO_K2n}.
	\item For $K_{1,n}$, there are labeled simple drawings which are CO-isomorphic, but not ERS-isomorphic (by relabeling the vertices in the larger partition class). 
\end{enumerate}

\subsubsection*{Outline.}

We prove Theorem~\ref{thm:orderRS} in Section~\ref{sec:RSCO_SI}, Section~\ref{sec:three} is devoted to proving Theorems~\ref{thm:rot}~and~\ref{thm:cr_ers}, and details of the non-implication results are given in Section~\ref{sec:counterexamples}. We conclude with an outlook on future work in Section~\ref{sec:conclusion}.\\
\emph{For all omitted or sketched proofs, full versions can be found in the appendix.}  

\section{Drawings of General Complete Multipartite Graphs}\label{sec:RSCO_SI}
While most of our results depend on the sizes of the partition classes of complete multipartite graphs, the following result holds for all complete multipartite graphs, independent of the sizes of their partition classes.
 
\orderRS*

We first state Theorem~\ref{thm:orderRS} for special complete bipartite graphs and small complete multipartite graphs in Lemma~\ref{lem:orient} and then use this result for proving the full theorem.
We sketch the proofs of Lemma~\ref{lem:orient} and Theorem~\ref{thm:orderRS}; their full proofs are deferred to Appendix~\ref{app:sec:RSCO_SI}.

\begin{restatable}{lemma}{lemorient}\label{lem:orient}
	Theorem~\ref{thm:orderRS} holds for $K_{1,n}$ and for complete multipartite graphs with at most four vertices (in the graph) and for~$K_{2,3}$.
\end{restatable}

\begin{figure}[htb]
	\centering
	\includegraphics[page=1]{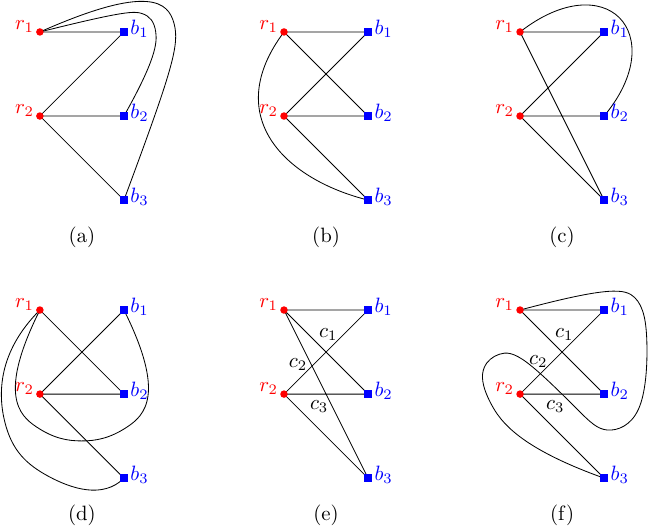}
	\caption {The (up to relabeling or strong isomorphism) only 6 drawings of~$K_{\textcolor{red}{2},\textcolor{blue}{3}}$, sorted by non-decreasing number of crossings.}
	\label{fig:K23}
\end{figure}

\begin{proof}[sketch]
For simple drawings of $K_{1,n}$ and $K_{1,1,1}=K_3$, the proof is trivial since the drawings are plane. 
For each complete multipartite graph on four vertices, there is, up to strong isomorphism and relabeling, only one unique drawing with a crossing and one unique drawing without a crossing. 
Hence, all unlabeled simple drawings of such graphs which are crossing CE-isomorphic are also strongly isomorphic and thus also all such drawings that are CO-isomorphic. 
There are, up to strong isomorphism and relabeling, only six drawings of $K_{2,3}$~\cite{harborth_enumeration}. 
They are depicted in  Figure~\ref{fig:K23}. The only two non strongly isomorphic unlabeled drawings that are CE-isomorphic are the ones in Figure~\ref{fig:K23}e and  Figure~\ref{fig:K23}f. They are not CO-isomorphic and we show that there is no labeling such that the drawings are CO-isomorphic. For each labeled drawing, there are twelve possible labelings, but we show that there is at most one relabeling of the original drawing such that the relabeled drawing is CO-isomorphic to the originally labeled drawing. For this pair of drawings we then show that they are ERS-isomorphic. 
Since ERS-isomorphism together with CO-isomorphism implies strong isomorphism by Theorem~\ref{thm:kyncl}, this concludes the proof of Lemma~\ref{lem:orient}.
\end{proof}

\begin{proof}[sketch of Theorem~\ref{thm:orderRS}]
	Let $D$ and $D'$ be two simple drawings of a complete multipartite graph $G$ and let $L_D$ and $L_{D'}$ be labelings of $D$ and $D'$, respectively, such that the thus labeled drawings are CE-isomorphic, RS-isomorphic, and CO-isomorphic.
	Consider a crossing in~$D$ and the subdrawing~$H$ induced by the four vertices involved by that crossing. Since $G$ is complete multipartite and thus $H$ is a drawing of a complete multipartite subgraph of $G$, there is a subdrawing $H_{2,2}$ of $H$ that is a simple drawing of $K_{2,2}$ and there is a subdrawing $H_{2,3}$ of $D$ that contains $H_{2,2}$ and is a simple drawing of $K_{2,3}$. (For $H_{2,2}$ the endvertices of the two crossing edges get split into the bipartition classes such that each edge still has two vertices of different bipartition classes; and for $H_{2,3}$ any arbitrary fifth vertex can be used. See Appendix~\ref{app:sec:RSCO_SI} for a rigorous proof.)
	
	We then compare $H_{2,3}$ in $D$ to its corresponding subdrawing $H'_{2,3}$ of $D'$, that is, the drawing induced by the same vertices and edges. Since $D$ and $D'$ are CO-isomorphic and RS-isomorphic with respect to the labeling $L(D)$ and $L(D')$ also $H_{2,3}$ and $H'_{2,3}$ are CO-isomorphic and RS-isomorphic and thus, by Lemma~\ref{lem:orient}, are strongly isomorphic. In particular, the ERS is the same or inverse. Thus, if the rotation system is the same in both drawings, so is the rotation of the crossing and analogously if the rotation system is inverse, so is the rotation of the crossing. Since this holds for all crossings of the drawings, $D$ and $D'$ have to be ERS-isomorphic.
Hence, $D$ and $D'$ are also ERS-isomorphic. Thus, by Theorem~\ref{thm:kyncl}, the drawings $D$ and $D'$ are strongly isomorphic.
\end{proof}

\section{Drawings of Complete Multipartite Graphs Where Each Partition Class Contains at Least Three Vertices}
\label{sec:three}

In this section, we consider complete multipartite graphs where each partition class contains at least three vertices. 
We will prove both Theorem~\ref{thm:rot} and Theorem~\ref{thm:cr_ers} via subdrawings of $K_{3,3}$ and in particular use the following result.

\begin{restatable}{lemma}{lemsmallk}\label{lem:weak_rot_small}
Any two labeled drawings of~$K_{3,3}$ that are CE-isomorphic are also RS-isomorphic. Moreover, if the two labeled drawings of~$K_{3,3}$ are CR-isomorphic in addition to RS-isomorphic, then the two drawings are ERS-isomorphic. 
\end{restatable}

\begin{proof}
Lemma~\ref{lem:weak_rot_small} is computer-assisted and  has been verified by considering all labeled drawings of $K_{3,3}$ and comparing the rotation systems of those drawings for which the crossing edge pairs are the same.
To this end, the enumeration of all 102 unlabeled simple drawings $K_{3,3}$ by Harborth~\cite{harborth_enumeration} has been encoded in a computer readable way. The enumeration of~\cite{harborth_enumeration} has also been independently shown in~\cite{bac_broetzner,bac_prinoth}.  

In more detail, for each of the 102 different unlabeled drawings of $K_{3,3}$ we read the encoded information about the rotation system and the crossing edges, including their rotation, from a given text file.
For each drawing we then generate all possible labelings (72 per drawing, $3!$ different labelings for each color class, times the exchange of red and blue).
This results in a set $S$ of a total of 7344 labeled drawings.
For all 26 963 496 pairs of drawings from $S$ we check if the two labeled drawings are CE-isomorphic by simply comparing their list of crossed edges.
Next, for the 10332 CE-isomorphic pairs we check if both drawings have the same or inverse rotation systems (which is straightforward as we have labeled drawings).
It turns out that this is in fact the case for all CE-isomorphic pairs.
So in total this implies that regardless which unlabeled two drawings of $K_{3,3}$ we take, and regardless how we label them,
if they have the same crossing edge pairs, then they have the same  or inverse rotation system, which implies the first part of Lemma~\ref{lem:weak_rot_small}.
Finally, for all CE-isomorphic pairs, we test if they are CR-isomorphic. For the resulting 4680 CR-isomorphic pairs, we verify that either the rotation system and the crossing rotations are both the same, or both are inverse. In other words, they are ERS-isomorphic.
The total running time of this routine is less than two seconds on a standard computer.

The program code, the input data, and a short description how to use and verify the correctness of the program are available online at~\cite{program_source}.
\end{proof}

Using Lemma~\ref{lem:weak_rot_small}, we first show Theorem~\ref{thm:rot} for bipartite graphs.

\begin{lemma}\label{thm:weak_rot_bipartite}
	Theorem~\ref{thm:rot} holds for~$K_{m,n}$ with $n \geq m \geq 3$.
\end{lemma}

\begin{proof}
	For simple drawings of $K_{\textcolor{red}{m},\textcolor{blue}{n}}$ with $\textcolor{red}{m}\geq 3, \textcolor{blue}{n} \geq 3$ we will show how to determine the rotation system from the information on crossing edge-pairs. 
	Let the vertex set 
	be split in a red set $\{\textcolor{red}{r_1},\ldots,\textcolor{red}{r_m}\}$ and a blue set $\{\textcolor{blue}{b_1},\ldots,\textcolor{blue}{b_n}\}$.
	We will determine the rotation system from the information of small subsets in three steps.
	In the first step, we find the rotation system of a subdrawing that is a~$K_{\textcolor{red}{3},\textcolor{blue}{3}}$ and includes vertices \textcolor{red}{$r_1$} and~\textcolor{blue}{$b_1$} by using Lemma~\ref{lem:weak_rot_small}. 
	This rotation system will determine 
	the global orientation of the rotation system.  
	In the second step, we determine the rotations around vertices \textcolor{red}{$r_1$} and~\textcolor{blue}{$b_1$}. We find them by using Lemma~\ref{lem:weak_rot_small} to sort incident edges in the rotation. In the third step, we determine the rotations around the remaining vertices.

	\noindent{\bf Step 1:} We first consider the subdrawing induced by~$\{\textcolor{red}{r_1,r_2,r_3},\textcolor{blue}{b_1,b_2,b_3}\}$ and find the rotation system of this subdrawing in the following way: It is a simple drawing of~$K_{\textcolor{red}{3},\textcolor{blue}{3}}$, so we know from Lemma~\ref{lem:weak_rot_small} that there are only two possible rotation systems, which are inverse to each other. There are only two possibilities for the rotation around~\textcolor{red}{$r_1$}: either~$\{\textcolor{blue}{b_1,b_2,b_3}\}$ or~$\{\textcolor{blue}{b_1,b_3,b_2}\}$. We choose the rotation such that~\textcolor{red}{$r_1$} has rotation~$\{\textcolor{blue}{b_1,b_2,b_3}\}$. (This will be the only choice for our proof, thus determining the rotation system. Choosing the inverse rotation would give the inverse rotation system.)
	
	\noindent{\bf Step 2:}  We now find the rotation around vertex~\textcolor{red}{$r_1$}: We look at subdrawings induced by~$\{\textcolor{red}{r_1,r_2,r_3},\textcolor{blue}{b_1,b_i,b_j}\}$. These are simple drawings of~$K_{\textcolor{red}{3},\textcolor{blue}{3}}$ so by Lemma~\ref{lem:weak_rot_small} there are only two possible rotation systems.  
	We have already fixed the rotation around~\textcolor{blue}{$b_1$}, thus the rotation system of these subgraphs is fixed. This way we learn in which order~\textcolor{blue}{$b_i$} and~\textcolor{blue}{$b_j$} are in the rotation around~\textcolor{red}{$r_1$} (using~\textcolor{blue}{$b_1$} as reference point). 	Doing this for different pairs~$\{\textcolor{blue}{b_i,b_j}\}$ we have a way to sort the vertices $\{\textcolor{blue}{b_1,b_2, ..., b_n}\}$ around~\textcolor{red}{$r_1$}, that is, to determine the rotation around~\textcolor{red}{$r_1$}. 
	
	We find the rotation around vertex~\textcolor{blue}{$b_1$} analogously 
	by considering the subdrawings induced by~$\{\textcolor{red}{r_1,r_i,r_j},\textcolor{blue}{b_1,b_2,b_3}\}$ and proceeding as described for~\textcolor{red}{$r_1$}. 

	\noindent{\bf Step 3:} Finally, we find the rotation around the remaining vertices in the following way: To find the rotation around vertex~\textcolor{red}{$r_i$} for $i>2$, we look at subdrawings induced by~$\{\textcolor{red}{r_1,r_2,r_i},\textcolor{blue}{b_1,b_j,b_k}\}$. These are simple drawings of~$K_{\textcolor{red}{3},\textcolor{blue}{3}}$ and so by Lemma~\ref{lem:weak_rot_small} the rotation system is determined up to inversion. As we already know the rotation around~\textcolor{blue}{$b_1$},  we can obtain the rotation system of this subgraph. As in Step~2, we sort the vertices in the rotation around~\textcolor{red}{$r_i$}, thus determining the rotation around~\textcolor{red}{$r_i$}. We repeat this process for all remaining vertices \textcolor{red}{$r_i$} for $i>2$ and for \textcolor{red}{$r_2$} we do the same process with $\{\textcolor{red}{r_1,r_2,r_3},\textcolor{blue}{b_1,b_j,b_k}\}$. Analogously, we find the rotation around all vertices~\textcolor{blue}{$b_i$}. 
\end{proof}

This result for complete bipartite graphs implies the statement for complete multipartite graphs.

\thmrot*

\begin{proof}
	Let $G$ be a complete multipartite graph in which each partition class has at least three vertices and let $D$ and $D'$ be two CE-isomorphic drawings of~$G$. Let $A$ be one of the partition classes. Then the subgraph $G_A$ of $G$ consisting of all vertices of $G$ and exactly the edges with one endpoint in $A$ is a complete bipartite graph where both bipartition classes contain at least three vertices. Thus, by Lemma~\ref{thm:weak_rot_bipartite}, the rotation in the drawing of $G_A$ that is a subdrawing of $D$ and the drawing of $G_A$ that is a subdrawing of $D'$ are either the same or inverse. Assume without loss of generality that they are the same (the case in which they are inverse is analogous). Since all edges incident to $A$ are in $G_A$, this determines all rotations of vertices in $A$ (to be the same in $D$ and $D'$). Let $B$ be a partition class of $G$ that is different from $A$. 
	Analogously to before, the two drawings of subgraph $G_B$ of $G$ consisting of all vertices of $G$ and exactly the edges with one endpoint in $B$ have the same or inverse rotation systems by Lemma~\ref{thm:weak_rot_bipartite}. $G_B$ and $G_A$ both contain the complete bipartite graph $G_{A,B}$ induced by the partition classes $A$ and $B$. Since we assumed that the drawings of $G_A$ have the same rotation system, the subdrawing induced by $G_{A,B}$ must have the same rotation system also in $G_B$. Thus, the drawing of $G_B$ that is a subdrawing of $D$ and the one that is a subdrawing of $D'$ have the same rotation system. This determines the rotations of all vertices in $B$ (to be the same rotation in~$D$ and~$D'$). Continuing analogously for each partition class, the rotations of all vertices are determined and $D$ and $D'$ must have the same rotation system. 
\end{proof}

\corcr*

\begin{proof} 
	Let $G$ be a complete multipartite graph in which each partition class has at least three vertices, let $D$ and $D'$ be two drawings with labelings 
	such that the two labeled drawings are CR-isomorphic drawings of~$G$. Then the labeled drawings are also CE-isomorphic and by Theorem~\ref{thm:rot} RS-isomorphic. 
	
	Assume, without loss of generality, that the rotation system in both drawings is the same (the case if they are inverse follows analogously, only with the drawings mirrored). If the rotations around all crossings are also the same in both drawings, then both drawings are ERS-isomorphic. So assume all crossings are inverse. Then also in every subdrawing of $D$ that is a $K_{3,3}$, all crossings are inverse while all rotations are the same, which contradicts Lemma~\ref{lem:weak_rot_small}.  
\end{proof}

Corollary~\ref{thm:semi} is a near-direct consequence of Theorems~\ref{thm:orderRS} and~\ref{thm:rot}. Any two CO-isomorphic drawings are trivially CE-isomorphic. Thus, by Theorem~\ref{thm:rot}, they are RS-isomorphic and consequently, by Theorem~\ref{thm:orderRS} be strongly isomorphic. 

\section{Examples of Simple Drawings Showing Specific Properties} 
\label{sec:counterexamples} 

\begin{figure}
	\centering
	\includegraphics[page=13]{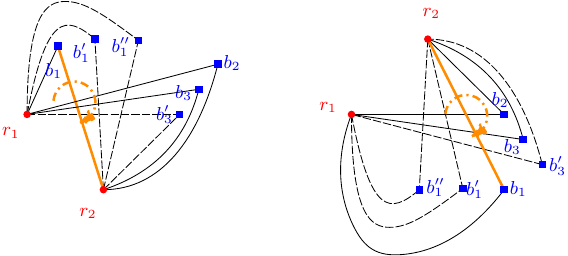}
	\caption {Two labeled simple drawings of~$K_{\textcolor{red}{2},\textcolor{blue}{3}}$, which are CR-isomorphic and RS-isomorphic, but not ERS-isomorphic. 
	}
	\label{fig:CR_RS_K2n}
\end{figure}

In this section, we look at relations where no implications hold. In particular, we depict the examples mentioned in the introduction. That is, for any combination of the characteristics \RS, \CEP, \CR, \CO, and \ERS for which there are simple drawings of complete multipartite graphs sharing exactly those, but no others, we give an example of such simple drawings. 
The depicted drawings are labeled. However, except for Figure~\ref{fig:CR_RS_K2n}, all statements also hold when seeing the drawings as unlabeled, that is, for every relabeling such that the isomorphisms between the relabeled drawings is a super-set of the isomorphism between the originally labeled drawings, the isomorphisms are actually the same; see Appendix~\ref{app:sec:counterexamples} for proofs of the unlabeled cases. The proofs for the labeled cases can be derived from the depicted drawings via their rotation and crossing properties. 
We list 
the proof details only for Figure~\ref{fig:onlyROT}, other proofs are similar; for details on the remaining proofs see Appendix~\ref{app:sec:counterexamples}.

In each of the drawings, we highlight edges that behave differently by drawing them bold and orange. Further, where relevant, we indicate that crossings have the same rotation in the compared drawings by drawing a solid, green arc around them and indicate that crossings have inverse rotations between the compared drawings by drawing an orange dashed-dotted arc around them. 
Finally, the vertices that have a dash in their label and are incident to dashed lines indicate how to extend the drawing by making arbitrary copies of that vertex, e.g. the $K_{\textcolor{red}{2},\textcolor{blue}{3}}$ drawn solid in Figure~\ref{fig:onlyROT} can be extended via~$\textcolor{blue}{b'}$ and copies of it to~$K_{\textcolor{red}{2},\textcolor{blue}{n}}$ for $\textcolor{blue}{n}\geq4$. 

\begin{figure}
		\centering
		\includegraphics[page=3]{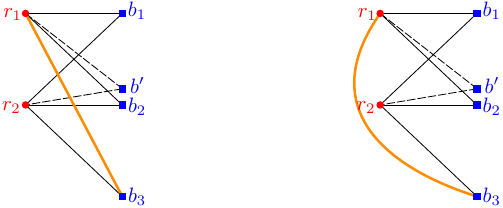}
	\caption {Simple drawings of~$K_{\textcolor{red}{2},\textcolor{blue}{3}}$ that are RS-isomorphic, but not CE-isomorphic. In particular, in both drawings, the rotations of both $\textcolor{red}{r_1}$ and $\textcolor{red}{r_2}$ are $\textcolor{blue}{b_1}$, $\textcolor{blue}{b_2}$, $\textcolor{blue}{b_3}$, and since the blue vertices have only degree two there is only one possible rotation for each of them. However, in the left drawing there exist crossings between the (bold, orange) edge $\textcolor{red}{r_1}\textcolor{blue}{b_3}$ and edges $\textcolor{red}{r_2}\textcolor{blue}{b_1}$ and $\textcolor{red}{r_2}\textcolor{blue}{b_2}$, while in the right drawing no edge crosses $\textcolor{red}{r_1}\textcolor{blue}{b_3}$.
    The drawings can be extended via $\textcolor{blue}{b'}$ (and the dashed edges incident to it) and copies of it to~$K_{\textcolor{red}{2},\textcolor{blue}{n}}$ for $\textcolor{blue}{n}\geq4$. 
	}
\label{fig:onlyROT}
\end{figure}

\newpage

\section{Conclusion}\label{sec:conclusion}
\begin{figure}
	\centering
	\includegraphics[page=8]{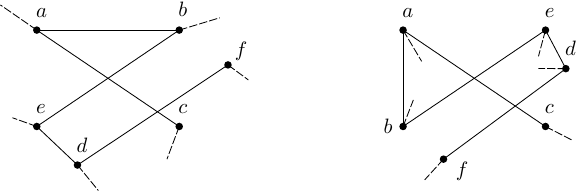}
	\caption {Two simple drawings of a connected graph on six vertices, which are CE-isomorphic, RS-isomorphic, and CO-isomophic, but not CR-isomorphic.
	}
	\label{fig:general_graph}
\end{figure}

From our results on implication between different types of isomorphism it follows that for drawings of complete multipartite graphs in which each partition class contains at least three vertices, there are four relevant classes of isomorphism: One class contains strong and CO-isomorphism; one contains CR- and ERS-isomorphism; one contains CE-isomorphism; and the last one contains RS-isomorphism. 
We remark that there are graphs for which less implications hold; see the labeled drawings in Figure~\ref{fig:general_graph} (this is extendable to  unlabeled drawings by vertices along the dashed lines).

In addition to relevantly improving the understanding of drawings of complete multipartite graphs, 
we believe that our characterization
can serve as a basis for further studying these graphs and drawings and obtaining new results on them.

It might also be helpful for research on simple drawings of other graphs.
In this context,  the question arises which other relevant graph classes admit or do not admit which kind of isomorphism implications.

\bibliography{multipartite_isomorphism_gd}

\newpage
\appendix
\section{Drawings of $K_{1,n}$}\label{sec:K1n}

\begin{figure} 
	\centering
	\includegraphics[scale=0.9,page=9]{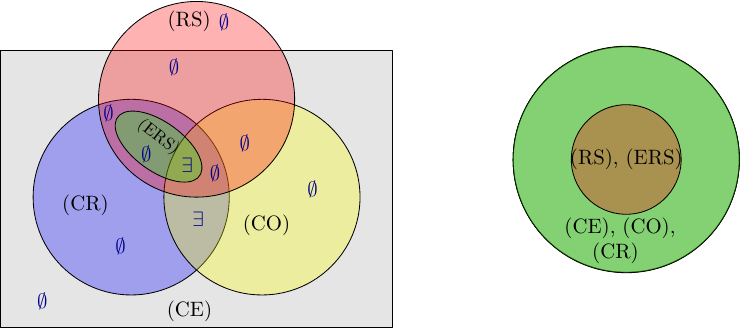}
	\caption {
		Classification of all possible combinations of different properties for labeled simple drawings of~$K_{1,n}$. For unlabeled drawings of~$K_{1,n}$, only the combination of all isomorphisms (located in the very middle) exists as all drawings are strongly isomorphic. 
	}
	\label{fig:venn_K1}
\end{figure}

Drawings of the $K_{1,n}$ behave differently than those of most other graphs; see Figure~\ref{fig:venn_K1} for a depiction. Since all edges are adjacent, no simple drawing of $K_{1,n}$ can have any crossings. Thus, all simple drawings of $K_{1,n}$ vacuously are CE-isomorphic, CR-isomorphic, and CO-isomorphic (as there are no crossings that can be different). Further, there is only one vertex with degree higher than one and thus only one vertex that can be assigned different rotations. Thus, any two unlabeled drawings of $K_{1,n}$ are strongly isomorphic. However, for $n\geq 3$, labeled drawings can have different rotation systems and thus different extended rotation systems. Trivially, if two labeled simple drawings of $K_{1,n}$ are RS-isomorphic (or ERS-isomorphic), then they are strongly isomorphic. 
Figure~\ref{fig:venn_K1}(left) displays all possible intersections, while the Figure~\ref{fig:venn_K1}(right) is drawn such that only areas of characteristics appear, for which there exists drawings of the considered graph that share those characteristics (and no others). 

\section{Missing Proofs of Section~\ref{sec:RSCO_SI}}\label{app:sec:RSCO_SI} 
We repeat the statements for convenience.
\lemorient*

\begin{figure}[htb]
\centering
\includegraphics[page=1]{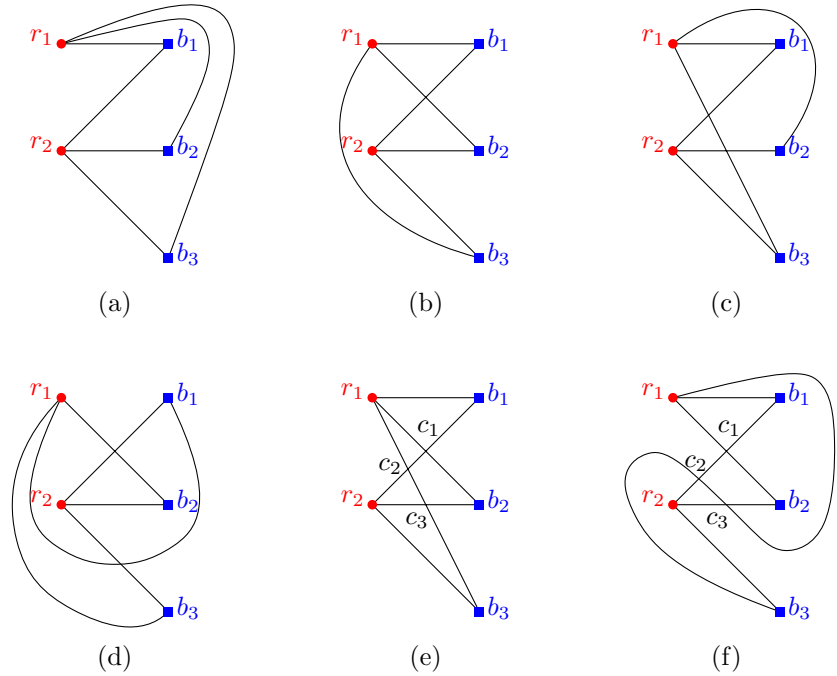}
\caption {The (up to relabeling or strong isomorphism) only 6 drawings of~$K_{2,3}$ sorted by non-decreasing number of crossings. Only drawings (e) and (f) are CE-isomorphic, but they are not CR-isomorphic. Drawings (b) and (e) are RS-isomorphic, but not CE-isomorphic. The same holds for drawings~(a),~(c), and~(d).
}
\label{app:fig:K23}
\end{figure}

\begin{proof}
Since all simple drawings of $K_{1,n}$ as well as all simple drawing of $K_{1,1,1}=K_3$ are plain, all drawings that are RS-isomorphic are strongly isomorphic. 

For each of $K_{2,2}$, $K_{1,1,2}$, and $K_{1,1,1,1}=K_4$, there is, up to strong isomorphism and relabeling, only one unique drawing with a crossing and one unique drawing without a crossing. 
Hence, all unlabeled simple drawings of $K_{2,2}$, $K_{1,1,2}$, and $K_{1,1,1,1}=K_4$ which are crossing CE-isomorphic are also strongly isomorphic and ´consequently CO-isomorphic. 
(For these small graphs, this only holds for unlabeled drawings and not labeled drawings, because in labeled drawings they might have one crossing with different rotation.)

There are, up to strong isomorphism and relabeling, only six drawings of~$K_{2,3}$~\cite{harborth_enumeration}. They are depicted in  Figure~\ref{app:fig:K23}. We show first that for any labeled drawing of~$K_{2,3}$, every relabeling that is CO-isomorphic and RS-isomorphic to the original drawing is also ERS-isomorphic -- and thus, by Theorem~\ref{thm:kyncl}, strongly isomorphic -- to the original drawing.  
We then show that for the only unlabeled drawings that are not strongly isomorphic, but CE-isomorphic --the ones in Figure~\ref{fig:K23}e and  Figure~\ref{app:fig:K23}f -- there cannot be a relabeling such that they are CO-isomorphic. 

We first observe that the drawings cannot be recolored as there are only two red vertices but three blue vertices. 
Thus, for each drawing, there are twelve possible relabelings - exchanging the labels of the two red points, and all possible permutations for the blue vertices. However, knowing that the obtained relabeling must be CO-isomorphic (and thus CE-isomorphic) to the original relabeling allows us to significantly reduce the number of labelings we truly have to check.

We start with the drawing in Figure~\ref{app:fig:K23}a. It is plane and thus any RS-isomorphic drawings are ERS-isomorphic. 

In the drawing of Figure~\ref{app:fig:K23}b, the unique crossing is between edges $\textcolor{red}{r_1}\textcolor{blue}{b_1}$ and~$\textcolor{red}{r_2}\textcolor{blue}{b_2}$. Thus, $\textcolor{blue}{b_3}$ cannot change its label (as it must not appear in a crossing) and when exchanging the labels of the red vertices or exchanging the labels of $\textcolor{blue}{b_1}$ and~$\textcolor{blue}{b_2}$, both changes have to be done. Consequently, the only possible label that is different from the original, but CE-isomorphic is to exchange the labels of $\textcolor{red}{r_1}$ and $\textcolor{red}{r_2}$ with each other and the labels of  $\textcolor{blue}{b_1}$ and $\textcolor{blue}{b_2}$ with each other. The obtained ERS of this relabeling is exactly the inverse of the original labeling.

In the drawing of Figure~\ref{app:fig:K23}c, the unique edge that crosses two edges is the edge $\textcolor{red}{r_1}\textcolor{blue}{b_3}$. Thus, in any CE-isomorphic drawing the labeling of those two vertices is fixed, which also fixes the labeling of $\textcolor{red}{r_2}$. Further, in any CO-isomorphic drawing, when following the edge $\textcolor{red}{r_1}\textcolor{blue}{b_3}$ from $\textcolor{red}{r_1}$ to $\textcolor{blue}{b_3}$, it first crosses $\textcolor{red}{r_2}\textcolor{blue}{b_1}$ and then~$\textcolor{red}{r_2}\textcolor{blue}{b_2}$, which fixes also the labeling of the remaining vertices. Thus, any labeling of the drawing that is CO-isomorphic to the original is also ERS-isomorphic.

In the drawing of Figure~\ref{app:fig:K23}d, the 
only completely uncrossed edges are  $\textcolor{red}{r_1}\textcolor{blue}{b_3}$ and $\textcolor{red}{r_2}\textcolor{blue}{b_2}$. Thus, in any CE-isomorphic drawing, the vertex $\textcolor{blue}{b_3}$ (that is not incident to any of those edges) has to keep the same label and if the labels of the red vertices are exchanged or the labels of $\textcolor{blue}{b_2}$ and $\textcolor{blue}{b_3}$ are exchanged, then both changes have to be made. This gives a unique (other than the original) relabeling such that the drawing CE-isomorphic, both exchanges have to be made. Thus, there is a unique relabeling that is different from the original but CE-isomorphic. In this relabeling, the ERS is  the same as in the original labeling. 

In the drawing of Figure~\ref{app:fig:K23}e, the 
unique completely uncrossed edges are $\textcolor{red}{r_1}\textcolor{blue}{b_1}$ and $\textcolor{red}{r_2}\textcolor{blue}{b_3}$. Thus, in any CE-isomorphic drawing, the vertex $\textcolor{blue}{b_2}$ (that is not incident to any of those edges) has to keep the same label and if the labels of the red vertices are exchanged or the labels of $\textcolor{blue}{b_1}$ and $\textcolor{blue}{b_3}$ are exchanged, then both exchanges have to be made. Thus, there is a unique relabeling that is different from the original but CE-isomorphic. In this relabeling, the ERS is the same as in the original labeling. We remark that the order of crossings along the edges stays the same.

Finally, in the drawing of Figure~\ref{app:fig:K23}f, again the unique completely uncrossed edges are $\textcolor{red}{r_1}\textcolor{blue}{b_1}$ and $\textcolor{red}{r_2}\textcolor{blue}{b_3}$. Thus, in any CE-isomorphic drawing, the vertex $\textcolor{blue}{b_2}$ (that is not incident to any of those edges) has to keep the same label and if the labels of the red vertices are exchanged or the labels of $\textcolor{blue}{b_1}$ and $\textcolor{blue}{b_3}$ are exchanged, then both exchanges have to be made. Thus, there is a unique relabeling that is different from the original but CE-isomorphic. The drawing with this relabeling is not CO-isomorphic to the drawing with the original labeling. We remark that the order of crossings along the edge $\textcolor{red}{r_1}\textcolor{blue}{b_3}$ and the crossing along the edge $\textcolor{red}{r_2}\textcolor{blue}{b_1}$ both change.

What remains to shown is that also two not strongly isomorphic drawings that are CO-isomorphic have to be ERS-isomorphic. We consider the only two not strongly isomorphic pairs of drawings that are CE-isomorphic, that is, the drawings in Figure~\ref{app:fig:K23}e and~\ref{app:fig:K23}f. According to the labeling in Figure~\ref{app:fig:K23}e and~\ref{app:fig:K23}f, while the crossings order of $\textcolor{red}{r_2}\textcolor{blue}{b_3}$ is the same in both drawings, the order of crossings along the edge $\textcolor{red}{r_1}\textcolor{blue}{b_3}$ is different in both drawings. Since we observed before that no CE-isomorphic relabeling of Figure~\ref{app:fig:K23}e changes the order of crossings and relabeling Figure~\ref{app:fig:K23}f can only change the crossing order of both edges  $\textcolor{red}{r_2}\textcolor{blue}{b_3}$ and $\textcolor{red}{r_1}\textcolor{blue}{b_3}$ simultaneously, the two drawings are never CO-isomorphic independent of the labeling.

Since ERS-isomorphism together with CO-isomorphism implies strong isomorphism by Theorem~\ref{thm:kyncl}, this concludes the proof of Lemma~\ref{lem:orient}.
\end{proof}

\orderRS*

\begin{proof}
Let $D$ and $D'$ be two simple drawings of a complete multipartite graph~$G$ and let $L_D$ and $L_{D'}$ be labelings for of $D$ and $D'$, respectively, such that the thus labeled drawings are CE-isomorphic, RS-isomorphic, and CO-isomorphic. 
Consider a crossing in~$D$ and the subdrawing~$H$ induced by the four vertices involved by that crossing. Since adjacent edges may not cross and for every edge the two endpoints are in different bipartition classes, $H$ is a simple drawing of $K_{2,2}$, $K_{1,1,2}$, or $K_{1,1,1,1}=K_4$. 

We choose a subdrawing $H_{2,2}$ of $H$ that still contains the considered crossing, but is a simple drawing of~$K_{2,2}$ in the following way.  If $H$ is a drawing of $K_{2,2}$, then $H_{2,2}$ is exactly $H$. If $H$ is a drawing of $K_{1,1,2}$, the edge $e$ between the two vertices that are in the partition classes of $H$ containing only one vertex, is adjacent to all other edges. Thus, $e$ is not involved in the crossing and taking $H$ without $e$ is a simple drawing of $K_{2,2}$ that still contains the crossing. We take $H$ without $e$ as $H_{2,2}$. 
If $H$ is a drawing of $K_{1,1,1,1}$, then for $H_{2,2}$ we take as one partition class exactly one vertex of each edge involved in the crossing. By that the other partition class automatically consists of the other vertex of each edge involved in the crossing. The crossing then still exists in the simple drawing of $K_{2,2}$ that is $H_{2,2}$. 
(We remark that any two vertices that are in different bipartition classes of $H_{2,2}$ are also in different partition classes of~$D$. The converse might not hold.)

Now, we obtain from $H_{2,2}$ a simple drawing $H_{2,3}$ that is a simple drawing of~$K_{2,3}$ which is a subdrawing of $D$. Let $v$ be an arbitrary vertex of $D$ that is not in $H_{2,2}$. Since all vertices that are in different bipartition classes of $H_{2,2}$ are also in different partition classes of~$D$, there has to be at least one bipartition class of $H_{2,2}$ containing vertices~$v_1,v_2$ such that in $D$, the vertex $v$ is in a different class than either of the two vertices~$v_1,v_2$. Thus, in $D$ there exists edges from $v$ to $v_1$ and to $v_2$. The subdrawing $H_{2,3}$ consists of $H_{2,2}$ together with the vertex $v$ and the edges from $v$ to $v_1$ and $v_2$.

Finally, we consider this subdrawing $H_{2,3}$ of $D$ and the corresponding subdrawing $H'_{2,3}$ of $D'$ that consists of the same edges and vertices as $H_{2,3}$. Both are drawings of the same $K_{2,3}$, and, by Lemma~\ref{lem:orient},  are strongly isomorphic. 
In particular, the ERS is the same or inverse. Thus, if the rotation system is the same in both drawings, so is the rotation of the crossing and analogously if the rotation system is inverse, so is the rotation of the crossing. Since this holds for all crossings of the drawings, $D$ and $D'$ have to be ERS-isomorphic. 

Thus, Properties~\ref{prop:order} and \ref{prop:rotation} of Theorem~\ref{thm:kyncl} are fulfilled and hence the drawings are strongly isomorphic.
\end{proof}

\section{Drawings of $K_{2,n}$}\label{sec:proof_K2n}

\begin{figure}[htb]
	\centering
	\includegraphics[page=10]{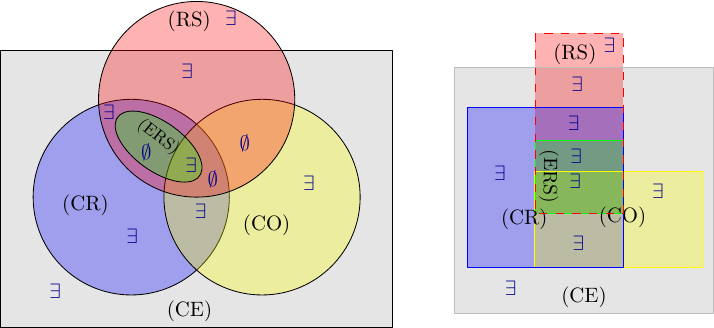}
	\caption {Classification of all possible combinations of different properties for labeled simple drawings of~$K_{2,n}$.}
	\label{fig:venn_K2}
\end{figure}

In this section, we consider simple drawings of $K_{2,n}$; see Figure~\ref{fig:venn_K2} for a depiction of the results. We give the proof for the implication of isomorphisms that holds additionally to Theorem~\ref{thm:orderRS}. We defer the proofs of which combinations of isomorphisms can exists between two simple drawings without implying other isomorphisms to Section~\ref{sec:counterexamples}.

\begin{restatable}{theorem}{thmsmall}\label{thm:K2n}
	Any two ERS-isomorphic simple drawings of~$K_{2,n}$ with  $n\geq 1$ are also strongly isomorphic.
\end{restatable}

We first show that Theorem~\ref{thm:K2n} holds for $n \leq 3$.

\begin{lemma}\label{lem:K2n}
	Let~$D$ and~$D'$ be two ERS-isomorphic simple drawings of~$K_{2,n}$, with $n \leq 3$, then~$D$ and~$D'$ are strongly isomorphic.
\end{lemma}

\begin{proof} We verify Lemma~\ref{lem:K2n} by looking at all simple drawings of those graphs (one drawing of $K_{2,1}$, two drawings of $K_{2,\textcolor{black}{2}}$, and six drawings of $K_{2,\textcolor{black}{3}}$~\cite{harborth_enumeration}) and checking that the extended rotation system is different in all of them. An easy way to see this is to follow along the lines of the arguments for Lemma~\ref{lem:orient}. 
	As argued there, all simple drawings of~$K_{1,\textcolor{black}{n}}$ that are RS-isomorphic are also strongly isomorphic. 
	Further, since any simple drawing of $K_{2,\textcolor{black}{2}}$ has at most one crossing and thus all such simple drawings are CO-isomorphic and consequently, by Theorem~\ref{thm:kyncl}, also strongly isomorphic. 
	There is only one pair of drawings of $K_{2,\textcolor{black}{3}}$ that has the same crossings, but is not strongly isomorphic. This pair consists of the drawings depicted in Figures~\ref{app:fig:K23}e and~\ref{app:fig:K23}f. In those two drawings, the rotation around some crossings -- crossings $c_2$ and $c_3$ according to the labeling of Figure~\ref{app:fig:K23} -- is different, while the rotation of other crossings-- $c_1$ according to the labeling of Figure~\ref{app:fig:K23} -- is the same. Thus, all drawings have different extended rotation systems. 
\end{proof}

With this result, we can prove Theorem~\ref{thm:K2n}. 

\begin{proof}[of Theorem~\ref{thm:K2n}]
	Let~$D$ and~$D'$ be labeled such that they are ERS-isomorphic. Since ERS-isomorphic also implies CE-isomorphic, by Theorem~\ref{thm:kyncl}, two ERS-isomorphic drawings that are also CO-isomorphic are also strongly isomorphic. So assume, for a contradiction, that there is an edge~$e=uv$ in~$D$ whose corresponding edge~$e'=u'v'$ in~$D'$ have the same crossings, but in different order. Then there is a pair of edges~$e_1,e_2$ in~$D$ and the corresponding pair~$e'_1, e'_2$ such that when going from~$u$ to~$v$ in~$D$, the edge~$e$ first crosses~$e_1$ and then~$e_2$, but when going from~$u'$ to~$v'$ in~$D'$, the edge~$e'$ first crosses~$e'_2$ and then~$e'_1$. The subdrawing~$H$ induced by~$e$,~$e_1$ and~$e_2$ is ERS-isomorphic to the subdrawing~$H'$ induced by~$e'$,~$e'_1$ and~$e'_2$. Since~$D$ is a simple drawing of~$K_{2,\textcolor{black}{n}}$~$e_1$ and~$e_2$ have to be incident. Thus~$H$ and consequently~$H'$ are simple drawings of~$K_{2,\textcolor{black}{3}}$. It follows from  Lemma~\ref{lem:K2n} that~$H$ and~$H'$ are strongly isomorphic, which is a contradiction.      
\end{proof}

\section{Examples of Simple Drawings Showing Specific Properties} 
\label{app:sec:counterexamples} 

In this section, we argue why the drawings presented in Section~\ref{sec:counterexamples} are indeed examples of drawings that are isomorphic with respect to exactly the types claimed there. For convenience, we repeat the drawings here, albeit sometimes with extended figure captions.  
 
\begin{figure}
\centering
\includegraphics[page=13]{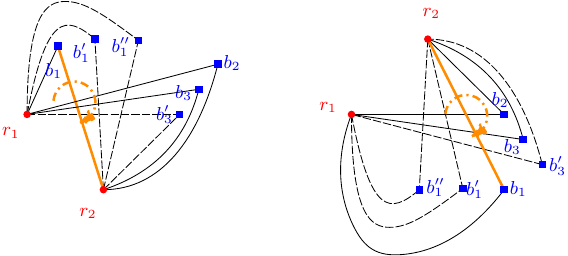}
\caption {Two labeled simple drawings of~$K_{\textcolor{red}{2},\textcolor{blue}{3}}$, which are CR-isomorphic and RS-isomorphic, but not ERS-isomorphic. 
	The rotation system is the same in both drawings. (The rotation of $\textcolor{red}{r_1}$ is  $\textcolor{blue}{b_1}$, $\textcolor{blue}{b_2}$, $\textcolor{blue}{b_3}$; and around it is $\textcolor{blue}{b_1}$, $\textcolor{blue}{b_3}$, $\textcolor{blue}{b_2}$. The blue vertices have only degree two, thus the rotations are the always the same.) However, the crossing rotations are inverse. The crossing edge pairs and their rotations are  $\textcolor{red}{r_1}\textcolor{blue}{b_2}$ crosses $\textcolor{red}{r_2}\textcolor{blue}{b_1}$ with rotation $\textcolor{red}{r_1}$, $\textcolor{blue}{b_1}$, $\textcolor{blue}{b_2}$, $\textcolor{red}{r_2}$ in the left drawing and with rotation  $\textcolor{red}{r_1}$, $\textcolor{red}{r_2}$ $\textcolor{blue}{b_2}$, $\textcolor{blue}{b_1}$ in the right drawing; and  $\textcolor{red}{r_1}\textcolor{blue}{b_3}$ crosses $\textcolor{red}{r_2}\textcolor{blue}{b_1}$ with rotation $\textcolor{red}{r_1}$, $\textcolor{blue}{b_1}$, $\textcolor{blue}{b_3}$, $\textcolor{red}{r_2}$ in the left drawing and with rotation  $\textcolor{red}{r_1}$, $\textcolor{red}{r_2}$ $\textcolor{blue}{b_3}$, $\textcolor{blue}{b_1}$ in the right drawing. We remark that also the crossing order along the (bold orange) edge $\textcolor{red}{r_2}\textcolor{blue}{b_1}$ is different.
	The drawings can be extended via $\textcolor{blue}{b_1'}$ or $\textcolor{blue}{b_3'}$ and copies of  one or both vertices (and the dashed edges incident to these vertices) to~$K_{\textcolor{red}{2},\textcolor{blue}{n}}$ for $\textcolor{blue}{n}\geq4$.
}
\label{app:fig:CR_RS_K2n}
\end{figure}

The figures we give are labeled and each figure caption contains the reasoning why the according statement holds for the labeled case. However, except for Figure~\ref{app:fig:CR_RS_K2n} (Figure~\ref{fig:CR_RS_K2n} in the main part), all statements also hold for unlabeled drawings. We show this by 
proving that the statements are independent of the labeling and, in particular, that there are are no labelings such that more characteristics are shared in addition to the claimed ones are (that is, assuming with the relabeling the two drawings are still isomorphic with all the types that the originally labels were isomorphic, then they cannot be also isomorphic in another sense).

\begin{figure} 
	\centering
	\includegraphics[page=3]{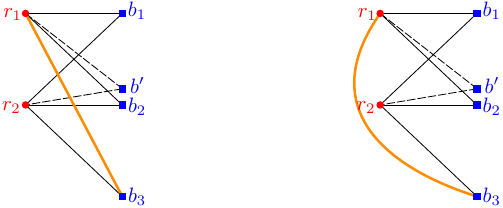}
	\caption {Simple drawings of~$K_{\textcolor{red}{2},\textcolor{blue}{3}}$ that are RS-isomorphic but different edge pairs cross 
		(they can be extended via $\textcolor{blue}{b'}$ and copies of it to~$K_{\textcolor{red}{2},\textcolor{blue}{n}}$ for $\textcolor{blue}{n}\geq4$). In particular, in both drawings, the rotation of $\textcolor{red}{r_1}$ and $\textcolor{red}{r_2}$ are both $\textcolor{blue}{b_1}$ $\textcolor{blue}{b_2}$ $\textcolor{blue}{b_3}$, and since the blue vertices have only degree two there is only one possible rotation for each of them. However, in the left drawing there exist crossings between the (bold, orange) edge $\textcolor{red}{r_2}\textcolor{blue}{b_1}$ and edges $\textcolor{red}{r_2}\textcolor{blue}{b_1}$, while in the right drawing no edge crosses $\textcolor{red}{r_2}\textcolor{blue}{b_1}$.
			The drawings can be extended via $\textcolor{blue}{b'}$ (and the dashed edges incident to it) and copies of it to~$K_{\textcolor{red}{2},\textcolor{blue}{n}}$ for $\textcolor{blue}{n}\geq4$.
	}
	\label{app:fig:onlyROT_small}
\end{figure}

\begin{figure} 
	\centering
	\includegraphics[page=10]{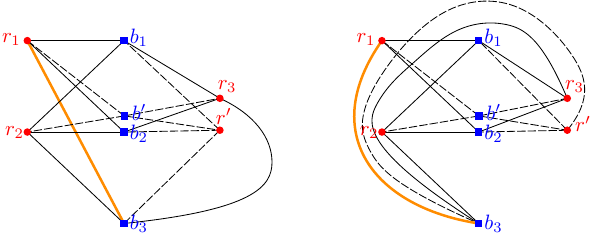}
	\caption {Simple drawings of~$K_{\textcolor{red}{4},\textcolor{blue}{3}}$ that are RS-isomorphic but different edge pairs cross. 
		(they can be extended via $\textcolor{blue}{b'}$, $\textcolor{red}{r'}$ and copies of them to~$K_{\textcolor{red}{m},\textcolor{blue}{n}}$ for $\textcolor{red}{m},\textcolor{blue}{n} \geq 4$). 
  The rotation system is $\textcolor{red}{r_1}$: $\textcolor{blue}{b_1}$, $\textcolor{blue}{b_2}$, $\textcolor{blue}{b_3}$; $\textcolor{red}{r_2}$: $\textcolor{blue}{b_1}$, $\textcolor{blue}{b_2}$, $\textcolor{blue}{b_3}$; $\textcolor{red}{r_3}$: $\textcolor{blue}{b_1}$, $\textcolor{blue}{b_3}$, $\textcolor{blue}{b_2}$; 
  	$\textcolor{blue}{b_1}$: $\textcolor{red}{r_1}$, $\textcolor{red}{r_3}$, $\textcolor{red}{r_2}$; $\textcolor{blue}{b_2}$: $\textcolor{red}{r_1}$, $\textcolor{red}{r_3}$, $\textcolor{red}{r_2}$;
  	$\textcolor{blue}{b_3}$: $\textcolor{red}{r_1}$, $\textcolor{red}{r_3}$, $\textcolor{red}{r_2}$. However, in the left drawing there exist crossings between the (bold, orange) edge $\textcolor{red}{r_1}\textcolor{blue}{b_3}$ and edges $\textcolor{red}{r_2}\textcolor{blue}{b_1}$ and $\textcolor{red}{r_2}\textcolor{blue}{b_2}$, while in the right drawing no edge crosses $\textcolor{red}{r_1}\textcolor{blue}{b_3}$.
  The drawings can be extended via $\textcolor{blue}{b'}$, $\textcolor{red}{r'}$ (and the dashed edges incident to these vertices) and copies of them to~$K_{\textcolor{red}{m},\textcolor{blue}{n}}$ for $\textcolor{blue}{m,n}\geq4$.	
}
	\label{app:fig:onlyROT_big}
\end{figure}

Figures~\ref{app:fig:onlyROT_small} and~\ref{app:fig:onlyROT_big} show simple drawings of complete multipartite graphs that are RS-isomorphic, but are not CE-isomorphic. 
In Figure~\ref{app:fig:onlyROT_small} the drawings are of  $K_{\textcolor{red}{2},\textcolor{blue}{3}}$ and can be extended to drawings of~$K_{\textcolor{red}{2},\textcolor{blue}{n}}$ by drawing vertices and edges like $\textcolor{blue}{b'}$ and the (incident) dashed edges. In Figure~\ref{app:fig:onlyROT_big}, a simple drawing of~$K_{\textcolor{red}{3},\textcolor{blue}{3}}$ is given, which can be extended to a simple drawing of $K_{\textcolor{red}{m},\textcolor{blue}{n}}$ with $\textcolor{blue}{n}\geq\textcolor{red}{m}\geq3$ as shown with $\textcolor{blue}{b'}$ and $\textcolor{red}{r'}$ and the dashed lines in Figure~\ref{app:fig:onlyROT_big}. For any $\textcolor{red}{m},\textcolor{blue}{n} \geq 3$, no 
relabeling can lead to CE-isomorphic drawings, because the number of completely uncrossed edges is different. 

\begin{figure} 
	\centering
	\includegraphics[page=7]{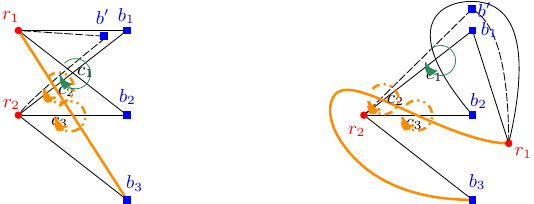}
	\caption {Two labeled simple drawings of~$K_{\textcolor{red}{2},\textcolor{blue}{3}}$ that are CE-isomorphic and RS-isomorphic, 
		but the crossings along the (bold, orange) edge $\textcolor{red}{r_1}$$\textcolor{blue}{r_3}$ are in different order and the rotations around crossings are different.
The rotation system is $\textcolor{red}{r_1}$: $\textcolor{blue}{b_1}$, $\textcolor{blue}{b_2}$, $\textcolor{blue}{b_3}$; $\textcolor{red}{r_2}$: $\textcolor{blue}{b_1}$, $\textcolor{blue}{b_2}$, $\textcolor{blue}{b_3}$. 
The crossing edge pairs are: $\textcolor{red}{r_1}\textcolor{blue}{b_2}$ crosses $\textcolor{red}{r_2}\textcolor{blue}{b_1}$ (we denote this crossing by $c_1$); $\textcolor{red}{r_1}\textcolor{blue}{b_3}$ crosses $\textcolor{red}{r_2}\textcolor{blue}{b_1}$ (we denote this crossing by $c_2$); and $\textcolor{red}{r_1}\textcolor{blue}{b_3}$ crosses $\textcolor{red}{r_2}\textcolor{blue}{b_2}$ (we denote this crossing by $c_3$). 
The rotation of the crossing $c_1$ (highlighted with a solid, green arc) is in both drawings $\textcolor{red}{r_1}$ $\textcolor{blue}{b_1}$ $\textcolor{blue}{b_2}$ $\textcolor{red}{r_2}$.
The rotation of the crossing $c_2$ (highlighted with a dash-dotted, orange arc) is inverse between the two drawings. In the left drawing it is $\textcolor{red}{r_1}$ $\textcolor{blue}{b_1}$ $\textcolor{blue}{b_3}$ $\textcolor{red}{r_2}$ and in the right drawing the rotation of crossing $c_2$ is $\textcolor{red}{r_1}$ $\textcolor{red}{r_2}$ $\textcolor{blue}{b_3}$ $\textcolor{blue}{b_2}$. 
When going along the bold, orange edge from $\textcolor{red}{r_1}$ to $\textcolor{blue}{b_3}$, in the left drawing crossing $c_2$ appears first and then crossing $c_3$, in the right drawing crossing $c_3$ appears before crossing $c_2$.
The drawings can be extended via $\textcolor{blue}{b'}$ and copies of it to~$K_{\textcolor{red}{2},\textcolor{blue}{n}}$ for $\textcolor{blue}{n}\geq4$.}
	\label{app:fig:CEP_RS_K2n_small}
\end{figure}

\begin{figure} 
	\centering
	\includegraphics[page=11]{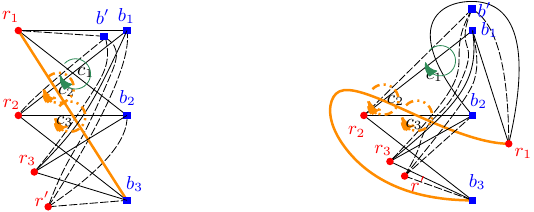}
	\caption {Two labeled simple drawings  of~$K_{\textcolor{red}{3},\textcolor{blue}{3}}$ that are CE-isomorphic and RS-isomorphic, 
		but the crossings along the (bold, orange) edge $\textcolor{red}{r_1}$$\textcolor{blue}{r_3}$ are in different order and the rotations of crossings is different.
		The rotation system is $\textcolor{red}{r_1}$: $\textcolor{blue}{b_1}$, $\textcolor{blue}{b_2}$, $\textcolor{blue}{b_3}$; $\textcolor{red}{r_2}$: $\textcolor{blue}{b_1}$, $\textcolor{blue}{b_2}$, $\textcolor{blue}{b_3}$; $\textcolor{red}{r_3}$: $\textcolor{blue}{b_1}$, $\textcolor{blue}{b_2}$, $\textcolor{blue}{b_3}$; $\textcolor{blue}{b_1}$: $\textcolor{red}{r_1}$, $\textcolor{red}{r_3}$, $\textcolor{red}{r_2}$; $\textcolor{blue}{b_2}$: $\textcolor{red}{r_1}$, $\textcolor{red}{r_3}$, $\textcolor{red}{r_2}$;	$\textcolor{blue}{b_3}$: $\textcolor{red}{r_1}$, $\textcolor{red}{r_3}$, $\textcolor{red}{r_2}$. 
		The crossing edge pairs are: 
		$\textcolor{red}{r_1}\textcolor{blue}{b_2}$ crosses $\textcolor{red}{r_2}\textcolor{blue}{b_1}$ (we denote this crossing by $c_1$); 
		$\textcolor{red}{r_1}\textcolor{blue}{b_2}$ crosses $\textcolor{red}{r_3}\textcolor{blue}{b_1}$; $\textcolor{red}{r_1}\textcolor{blue}{b_3}$ crosses $\textcolor{red}{r_2}\textcolor{blue}{b_1}$ (we denote this crossing by $c_2$); 
		$\textcolor{red}{r_1}\textcolor{blue}{b_3}$ crosses $\textcolor{red}{r_2}\textcolor{blue}{b_2}$ (we denote this crossing by $c_3$);
		$\textcolor{red}{r_1}\textcolor{blue}{b_3}$ crosses $\textcolor{red}{r_3}\textcolor{blue}{b_1}$;
		$\textcolor{red}{r_1}\textcolor{blue}{b_3}$ crosses $\textcolor{red}{r_3}\textcolor{blue}{b_2}$;
		$\textcolor{red}{r_2}\textcolor{blue}{b_2}$ crosses $\textcolor{red}{r_3}\textcolor{blue}{b_1}$;		
		$\textcolor{red}{r_2}\textcolor{blue}{b_3}$ crosses $\textcolor{red}{r_3}\textcolor{blue}{b_1}$;
		and $\textcolor{red}{r_2}\textcolor{blue}{b_3}$ crosses $\textcolor{red}{r_3}\textcolor{blue}{b_2}$. 	
		The rotation of the crossing $c_1$ (highlighted with a solid, green arc) is in both drawings $\textcolor{red}{r_1}$ $\textcolor{blue}{b_1}$ $\textcolor{blue}{b_2}$ $\textcolor{red}{r_2}$, but the rotation of the crossing $c_2$ (highlighted with a dash-dotted, orange arc) is inverse between the two drawings. In the left drawing it is $\textcolor{red}{r_1}$ $\textcolor{blue}{b_1}$ $\textcolor{blue}{b_3}$ $\textcolor{red}{r_2}$ and in the right drawing the rotation of crossing $c_2$ is $\textcolor{red}{r_1}$ $\textcolor{red}{r_2}$ $\textcolor{blue}{b_3}$ $\textcolor{blue}{b_2}$.
		When going along the bold, orange edge from $\textcolor{red}{r_1}$ to $\textcolor{blue}{b_3}$, in the left drawing crossing $c_2$ appears first and then crossing $c_3$, in the right drawing crossing $c_3$ appears before crossing $c_2$.
  The drawings can be extended via $\textcolor{blue}{b'}$, $\textcolor{red}{r'}$ (and the dashed edges incident to these vertices) and copies of them to~$K_{\textcolor{red}{m},\textcolor{blue}{n}}$ for $\textcolor{blue}{m,n}\geq4$.	
}
	\label{app:fig:CEP_RS_K2n_big}
\end{figure}

The drawings in Figures~\ref{app:fig:CEP_RS_K2n_small} and~\ref{app:fig:CEP_RS_K2n_big} show labeled simple drawings that are CE-isomorphic and RS-isomorphic, but the crossings are in a different order and don't have the same rotations. In Figure~\ref{app:fig:CEP_RS_K2n_small}, the underlying graph is a $K_{\textcolor{red}{2},\textcolor{blue}{3}}$, in Figure~\ref{app:fig:CEP_RS_K2n_big} the underlying graph is a $K_{\textcolor{red}{3},\textcolor{blue}{3}}$. There is one unique crossing ($c_2$) which is the crossing between two edges that are involved in two crossings each. In the left drawing, when following the involved edges from the vertex on the smaller (red) bipartition class to the other vertex, this unique crossing comes first for both edges. In the right drawing this does not hold. Thus, relabeling cannot result in the same crossing order. Further, this special crossing has a different rotation in the left drawing than it has in the right drawing. Hence, relabeling also cannot make the drawings have the same crossing rotations. These drawings can be extended to any $K_{\textcolor{red}{2},\textcolor{blue}{n}}$ with $\textcolor{blue}{n}\geq 3$ by making copies of vertex $\textcolor{blue}{b_1}$ as shown with $\textcolor{blue}{b'}$ and $\textcolor{blue}{b''}$ and the dashed edges in Figure~\ref{app:fig:CEP_RS_K2n_big}.

\begin{figure}
	\centering
	\includegraphics[page=9]{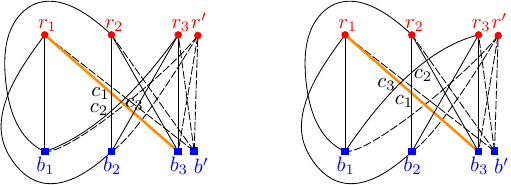}
	\caption {Two simple drawings of~$K_{\textcolor{red}{3},\textcolor{blue}{3}}$, which are ERS-isomorphic, but the crossings are in a different order.
	The rotation system is $\textcolor{red}{r_1}$: $\textcolor{blue}{b_1}$, $\textcolor{blue}{b_2}$, $\textcolor{blue}{b_3}$; $\textcolor{red}{r_2}$: $\textcolor{blue}{b_1}$, $\textcolor{blue}{b_3}$, $\textcolor{blue}{b_2}$; $\textcolor{red}{r_3}$: $\textcolor{blue}{b_1}$, $\textcolor{blue}{b_3}$, $\textcolor{blue}{b_2}$; $\textcolor{blue}{b_1}$: $\textcolor{red}{r_1}$, $\textcolor{red}{r_3}$, $\textcolor{red}{r_2}$; $\textcolor{blue}{b_2}$: $\textcolor{red}{r_1}$, $\textcolor{red}{r_2}$, $\textcolor{red}{r_3}$;	$\textcolor{blue}{b_3}$: $\textcolor{red}{r_1}$, $\textcolor{red}{r_2}$, $\textcolor{red}{r_3}$. 
	The crossing edge pairs and their rotations are: 
	$\textcolor{red}{r_1}\textcolor{blue}{b_2}$ crosses $\textcolor{red}{r_2}\textcolor{blue}{b_1}$ with rotation $\textcolor{red}{r_1}$, $\textcolor{blue}{b_1}$, $\textcolor{blue}{b_2}\textcolor{red}{r_1}$ ; 
	$\textcolor{red}{r_1}\textcolor{blue}{b_3}$ crosses $\textcolor{red}{r_2}\textcolor{blue}{b_2}$ (we denote this crossing by $c_1$) with rotation $\textcolor{red}{r_1}$, $\textcolor{red}{r_2}$, $\textcolor{blue}{b_3}$, $\textcolor{blue}{b_2}$; 
	$\textcolor{red}{r_1}\textcolor{blue}{b_3}$ crosses $\textcolor{red}{r_3}\textcolor{blue}{b_1}$ (we denote this crossing by $c_3$) with rotation $\textcolor{red}{r_1}$, $\textcolor{red}{r_3}$, $\textcolor{blue}{b_3}$, $\textcolor{blue}{b_1}$;
	$\textcolor{red}{r_1}\textcolor{blue}{b_3}$ crosses $\textcolor{red}{r_3}\textcolor{blue}{b_2}$ with rotation $\textcolor{red}{r_1}$, $\textcolor{red}{r_3}$, $\textcolor{blue}{b_3}$, $\textcolor{blue}{b_2}$; 
	$\textcolor{red}{r_2}\textcolor{blue}{b_2}$ crosses $\textcolor{red}{r_3}\textcolor{blue}{b_1}$ with rotation $\textcolor{red}{r_2}$, $\textcolor{red}{r_3}$, $\textcolor{blue}{b_2}$, $\textcolor{blue}{b_1}$; 		
	$\textcolor{red}{r_2}\textcolor{blue}{b_3}$ crosses $\textcolor{red}{r_3}\textcolor{blue}{b_1}$ with rotation $\textcolor{red}{r_2}$, $\textcolor{red}{r_3}$, $\textcolor{blue}{b_3}$, $\textcolor{blue}{b_1}$; 	
	and $\textcolor{red}{r_2}\textcolor{blue}{b_3}$ crosses $\textcolor{red}{r_3}\textcolor{blue}{b_2}$ with rotation $\textcolor{red}{r_2}$, $\textcolor{red}{r_3}$, $\textcolor{blue}{b_3}$, $\textcolor{blue}{b_2}$. 	
	When going along the bold, orange edge from $\textcolor{red}{r_1}$ to $\textcolor{blue}{b_3}$, in the left drawing crossing $c_1$ appears first and then crossing $c_3$, in the right drawing crossing $c_3$ appears before crossing $c_2$.
	The drawings can be extended via $\textcolor{blue}{b'}$, $\textcolor{red}{r'}$ (and the dashed edges incident to these vertices) and copies of them to~$K_{\textcolor{red}{m},\textcolor{blue}{n}}$ for $\textcolor{blue}{m,n}\geq4$.		
}
	\label{app:fig:ERS_Kmn}
\end{figure}

The drawings in Figure~\ref{app:fig:ERS_Kmn} show labeled simple drawings of $K_{\textcolor{red}{3},\textcolor{blue}{3}}$ that are ERS-isomorphic, but the order of crossings is different. There also cannot be a labeling of those crossings such that they have the same order because of the following reasons. There are only two edges that are involved with three crossings each ($\textcolor{red}{r_1}\textcolor{blue}{b_3}$ and $\textcolor{red}{r_3}\textcolor{blue}{b_1}$) that cross in $c_3$. According to the crossing order of the right drawing, $c_3$ is the first crossing of $\textcolor{red}{r_1}\textcolor{blue}{b_3}$ when going from $\textcolor{red}{r_1}$ to $\textcolor{blue}{b_3}$. According to the crossing order of the left drawing,  $c_3$ is always the second crossing for both edges (independent of the direction in which the edge is followed). 
Theses drawings can be extended to any $K_{\textcolor{blue}{m},\textcolor{red}{n}}$ with $m\geq 3$ and $n \geq 3$ by making copies of vertices $\textcolor{blue}{b_3}$ and $\textcolor{red}{r_3}$ as shown with $\textcolor{blue}{b'}$ and $\textcolor{red}{r'}$, respectively, and the dashed edges in Figure~\ref{app:fig:ERS_Kmn}.

\begin{figure}[hbt]
	\centering
	\includegraphics[page=1]{K2n_examples}
	\caption {Two simple drawings of~$K_{\textcolor{red}{2},\textcolor{blue}{5}}$, which are CE-isomorphic, but not RS-isomorphic or CO-isomorphic or CR-isomorphic. 
		The crossing edge pairs are: 
		$\textcolor{red}{r_1}\textcolor{blue}{b_2}$ crosses $\textcolor{red}{r_2}\textcolor{blue}{b_1}$ (we denote this crossing by $c_1$); 
		$\textcolor{red}{r_1}\textcolor{blue}{b_4}$ crosses $\textcolor{red}{r_2}\textcolor{blue}{b_3}$ (we denote this crossing by $c_2$);
		$\textcolor{red}{r_1}\textcolor{blue}{b_5}$ crosses $\textcolor{red}{r_2}\textcolor{blue}{b_4}$;
		and $\textcolor{red}{r_1}\textcolor{blue}{b_5}$ crosses $\textcolor{red}{r_2}\textcolor{blue}{b_3}$;
		$\textcolor{blue}{b_2}$.
		In the left picture, $\textcolor{blue}{b_3}$ and $\textcolor{blue}{b_5}$ appear consecutive in the rotation of $\textcolor{red}{r_1}$, but in the right picture, they are not consecutive, thus the rotation systems are neither the same nor inverse of each other. 
		The rotation of crossing $c_1$ is inverse between the two drawings (in the left drawing $\textcolor{red}{r_1}$, $\textcolor{blue}{b_2}$, $\textcolor{blue}{b_2}$, $\textcolor{red}{r_2}$; and in the right drawing $\textcolor{red}{r_1}$, $\textcolor{red}{r_2}$, $\textcolor{blue}{b_2}$, $\textcolor{blue}{b_1}$). However, the rotation of crossing $c_2$ is the same in both drawings ($\textcolor{red}{r_1}$, $\textcolor{blue}{b_3}$, $\textcolor{blue}{b_4}$, $\textcolor{red}{r_2}$). 
		When going along the bold, orange edge from $\textcolor{red}{r_1}$ to $\textcolor{blue}{b_5}$, in the left drawing crossing $c_4$ appears first and then crossing $c_3$, in the right drawing crossing $c_3$ appears before crossing $c_4$.
		The drawings can be extended via $\textcolor{blue}{b'}$ and copies of it (and the dashed edges incident to these vertices) to~$K_{\textcolor{red}{2},\textcolor{blue}{n}}$ for $\textcolor{blue}{n}\geq6$.
	}
	\label{app:fig:only_CEP_K2n}
\end{figure}

The drawings in Figure~\ref{app:fig:only_CEP_K2n} show simple drawings of $K_{\textcolor{red}{2},\textcolor{blue}{5}}$ that are CE-isomorphic, but not RS-isomorphic or CO-isomorphic or CR-isomorphic. In the following, we reason that all these properties hold for any relabeling. 
The number of vertices in the different bipartition classes is different, so the coloring cannot be exchanged. Further, in the right drawing, $\textcolor{red}{r_1}$ and $\textcolor{red}{r_2}$ are incident to two completely uncrossed edges that are consecutive in the rotation around the vertices, while in the left drawing the two uncrossed edges incident to $\textcolor{red}{r_1}$ are not consecutive. Hence, relabeling cannot lead to the same rotation system while keeping the same pairs of edges crossing. As the rotation around~$c_1$, the crossing between $\textcolor{red}{r_1}\textcolor{blue}{b_2}$ and $\textcolor{red}{r_2}\textcolor{blue}{b_1}$, is different in the two drawings, but the rotation around~$c_2$, the crossing between $\textcolor{red}{r_1}\textcolor{blue}{b_4}$ and $\textcolor{red}{r_2}\textcolor{blue}{b_3}$, is the same in the two drawings, relabeling also cannot make the crossing rotations the same. 
Further, consider all edges oriented from left to blue. Then in the right drawing both edges that cross two other edges first cross each other and then cross another edge, while in the left drawing, there is an edge that first crosses an edge that has only one crossing and then crosses the other edge which has two crossings. Thus, relabeling also cannot lead to the same crossing orders. The example can be generalized to a simple drawing of $K_{\textcolor{red}{2},\textcolor{blue}{n}}$ for any $\textcolor{red}{n}\geq 5$ by adding blue vertices to the unbounded face like $\textcolor{blue}{b'}$ (or copies of $\textcolor{blue}{b'}$) and edges between the new vertex to $\textcolor{red}{r_1}$ and $\textcolor{red}{r_2}$ (as the dashed lines in Figure~\ref{app:fig:only_CEP_K2n}) such that they are completely uncrossed.

\begin{figure}
\centering
\includegraphics[page=5]{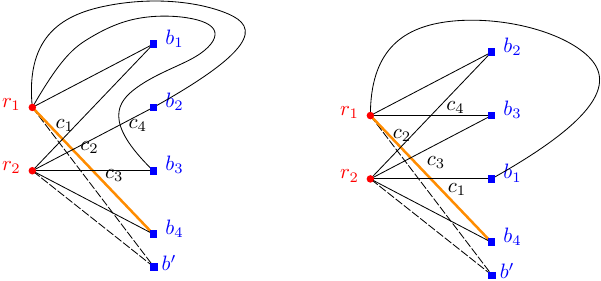}
\caption {Two simple drawings of~$K_{\textcolor{red}{2},\textcolor{blue}{4}}$, which are CR-isomorphic, but not RS-isomorphic or CO-isomorphic. 
The crossing rotations are in both drawings $c_1$: $\textcolor{red}{r_1}$, $\textcolor{blue}{b_1}$, $\textcolor{blue}{b_4}$, $\textcolor{red}{r_2}$; $c_2$: $\textcolor{red}{r_1}$, $\textcolor{blue}{b_2}$, $\textcolor{blue}{b_4}$, $\textcolor{red}{r_2}$; $c_3$: $\textcolor{red}{r_1}$, $\textcolor{blue}{b_3}$, $\textcolor{blue}{b_4}$, $\textcolor{red}{r_2}$; $c_4$: $\textcolor{red}{r_1}$, $\textcolor{blue}{b_2}$, $\textcolor{blue}{b_3}$, $\textcolor{red}{r_2}$. In the left drawing, $\textcolor{blue}{b_1}$ and $\textcolor{blue}{b_3}$ appear consecutive in the rotation of $\textcolor{red}{r_1}$, but in the right drawing they do not. Thus, the rotation system cannot be the same or inverse. When going along the bold, orange edge from $\textcolor{red}{r_1}$ to $\textcolor{blue}{b_4}$, in the left drawing crossing $c_1$ appears first and then crossing $c_2$ and $c_3$, in the right drawing crossings $c_2$ and $c_3$ appear before crossing $c_1$.
	The drawings can be extended via $\textcolor{blue}{b'}$ and copies of it (and the dashed edges incident to these vertices) to~$K_{\textcolor{red}{2},\textcolor{blue}{n}}$ for $\textcolor{blue}{n}\geq5$.
}
\label{app:fig:CEP_CR_K2n}
\end{figure}

The drawings in Figure~\ref{app:fig:CEP_CR_K2n} show labeled simple drawings of $K_{\textcolor{red}{2},\textcolor{blue}{4}}$ that are CR-isomorphic (and thus CE-isomorphic), but not RS-isomorphic or CO-isomorphic. There cannot be a labeling of those drawings such that they are CE-isomorphic and RS-isomorphic or a labeling such that they are CO-isomorphic for the following reasons.
There is one unique edge involved in three crossings (labeled~$\textcolor{red}{r_1}\textcolor{blue}{b_4}$), one unique edge involved in two crossings (labeled $\textcolor{red}{r_2}\textcolor{blue}{b_2}$), and one unique vertex that is on the smaller bipartition class and incident to the edge crossing three other edges. Of the blue vertices not incident to the edge crossing twice or the edge crossing three times, one is incident to an edge that crosses nothing~($\textcolor{red}{r_1}\textcolor{blue}{b_1}$) and the other vertex ($\textcolor{blue}{b_3}$) is not. Thus, in order to be CE-isomorphic, the labeling is fixed and consequently no relabeling can lead to the same crossing order or the same rotation system while still having the same crossing edge-pairs. Theses drawings can be extended to any $K_{\textcolor{red}{2},\textcolor{blue}{n}}$ with $\textcolor{blue}{n}\geq 3$ by making copies of vertex $\textcolor{blue}{b_4}$ as shown with $\textcolor{blue}{b'}$ and the dashed edges in Figure~\ref{app:fig:CEP_CR_K2n}.

The drawings in Figure~\ref{app:fig:CR_RS_K2n} show labeled simple drawings of $K_{\textcolor{red}{2},\textcolor{blue}{3}}$, which are CR-isomorphic and RS-isomorphic, but not ERS-isomorphic. The drawings can be extended to simple drawings of $K_{\textcolor{red}{m},\textcolor{blue}{n}}$ by copying the vertex $\textcolor{blue}{b_1}$ as shown with $\textcolor{blue}{b'_1}$ and $\textcolor{blue}{b''_1}$  or copying the vertex $\textcolor{blue}{b_3}$ as shown with $\textcolor{blue}{b'_3}$ (both can be copied an arbitrary number of times).
We remark that the unlabeled drawings in Figure~\ref{app:fig:CR_RS_K2n} are strongly isomorphic.

\begin{figure}
\centering
\includegraphics[page=4]{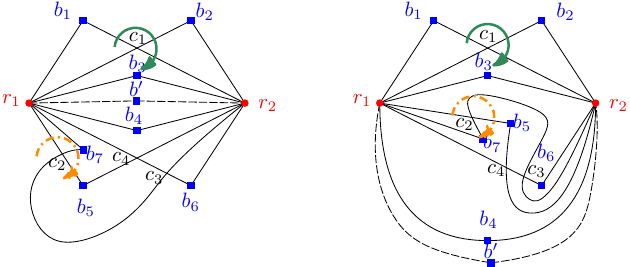}
\caption {Two simple drawings of~$K_{\textcolor{red}{2},\textcolor{blue}{7}}$, which are CE-isomorphic and CO-isomorphic, but neither RS-isomorphic nor CR-isomorphic.
	The crossing edge pairs are: 
	$\textcolor{red}{r_1}\textcolor{blue}{b_2}$ crosses $\textcolor{red}{r_2}\textcolor{blue}{b_1}$ (we denote this crossing by $c_1$); 
	$\textcolor{red}{r_1}\textcolor{blue}{b_6}$ crosses $\textcolor{red}{r_2}\textcolor{blue}{b_7}$ (we denote this crossing by $c_3$); 
	$\textcolor{red}{r_1}\textcolor{blue}{b_6}$ crosses $\textcolor{red}{r_2}\textcolor{blue}{b_5}$ (we denote this crossing by $c_4$);
	$\textcolor{red}{r_1}\textcolor{blue}{b_5}$ crosses $\textcolor{red}{r_2}\textcolor{blue}{b_7}$ (we denote this crossing by $c_2$). 
	The crossing order of the edges that have at least two crossings is: When going along the edge $\textcolor{red}{r_1}\textcolor{blue}{b_6}$ from $\textcolor{red}{r_1}$ to $\textcolor{blue}{b_6}$, first the crossing $c_4$ appears and then the crossing $c_3$; when going along the edge $\textcolor{red}{r_2}\textcolor{blue}{b_7}$ from $\textcolor{red}{r_2}$ to $\textcolor{blue}{b_7}$, first the crossing $c_3$ appears and then the crossing $c_2$.
	In the left drawing, $\textcolor{blue}{b_1}$ and $\textcolor{blue}{b_5}$ appear consecutive and in the right drawing, they do not, thus the rotation system cannot be the same or inverse. 
	The rotation of crossing $c_1$ is the same in both drawings ($\textcolor{red}{r_1}$, $\textcolor{blue}{b_1}$, $\textcolor{blue}{b_2}$, $\textcolor{red}{r_2}$). However, the rotation of crossing $c_2$ is inverse between the two drawings (in the left drawing $\textcolor{red}{r_1}$, $\textcolor{blue}{b_7}$, $\textcolor{blue}{b_5}$, $\textcolor{red}{r_2}$; and in the right drawing $\textcolor{red}{r_1}$, $\textcolor{red}{r_2}$, $\textcolor{blue}{b_5}$, $\textcolor{blue}{b_7}$).
	The drawings can be extended via $\textcolor{blue}{b'}$ and copies of it (and the dashed edges incident to these vertices) to~$K_{\textcolor{red}{2},\textcolor{blue}{n}}$ for $\textcolor{blue}{n}\geq8$.
}
\label{app:fig:CEP_CO_K2n}
\end{figure}

The drawings in Figure~\ref{app:fig:CEP_CO_K2n} show labeled simple drawings of $K_{\textcolor{red}{2},\textcolor{blue}{4}}$ that are CO-isomorphic (and thus CE-isomorphic), but neither CR-isomorphic nor RS-isomorphic. There cannot be a labeling in which they are RS-isomorphic and CE-isomorphic nor a labeling such that they are CR-isomorphic for the following reasons.
The number of vertices on the different sides of the bipartition are different. Thus, the coloring is fixed. There are exactly two blue vertices that are incident to two completely uncrossed edges ($\textcolor{blue}{b_3}$ and $\textcolor{blue}{b_4}$ with the currently labeling). In the left drawing these two are consecutive in the rotation around the red points, in the right drawing they are not. Thus, also after relabeling, these drawings cannot be RS-isomorphic and CE-isomorphic. Further, the unique crossing of two edges whose only crossing is with each other ($\textcolor{red}{r_2}\textcolor{blue}{b_1}$ and $\textcolor{red}{r_1}\textcolor{blue}{b_2}$) has the same crossing rotation, while the other crossings have inverse crossing rotations. Thus, it is not possible to relabel the vertices such that the drawings are CR-isomorphic. These drawings can be extended to any $K_{\textcolor{red}{2},\textcolor{blue}{n}}$ with $n\geq 3$ by making copies of vertex $\textcolor{blue}{b_4}$ as shown with $\textcolor{blue}{b'}$ and the dashed edges in Figure~\ref{app:fig:CEP_CO_K2n}.

\begin{figure}
\centering
\includegraphics[page=6]{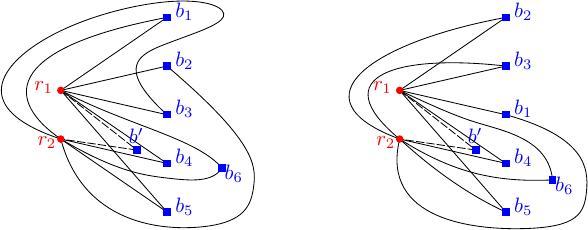}
\caption {Two simple drawings of~$K_{\textcolor{red}{2},\textcolor{blue}{6}}$, which are CO-isomorphic and CR-isomorphic, but not RS-isomorphic. In the left drawing, $\textcolor{blue}{b_3}$ and $\textcolor{blue}{b_6}$ are consecutive in the rotation around $\textcolor{red}{r_1}$ and in the right drawing they are not, thus the rotation systems are neither the same nor inverse of each other. The crossing edge pairs and their rotations are in both drawings: $\textcolor{red}{r_1}\textcolor{blue}{b_2}$ crosses $\textcolor{red}{r_2}\textcolor{blue}{b_3}$ with rotation $\textcolor{red}{r_1}$, $\textcolor{red}{r_2}$, $\textcolor{blue}{b_2}$, $\textcolor{blue}{b_3}$; the edge $\textcolor{red}{r_1}\textcolor{blue}{b_5}$ crosses~$\textcolor{red}{r_2}\textcolor{blue}{b_4}$ with rotation $\textcolor{red}{r_1}$, $\textcolor{blue}{b_4}$, $\textcolor{blue}{b_5}$, $\textcolor{red}{r_2}$; and $\textcolor{red}{r_1}\textcolor{blue}{b_5}$ crosses $\textcolor{red}{r_2}\textcolor{blue}{b_6}$ with rotation $\textcolor{red}{r_1}$, $\textcolor{blue}{b_6}$, $\textcolor{blue}{b_5}$, $\textcolor{red}{r_2}$. The only edge involved in two crossings is $\textcolor{red}{r_1}\textcolor{blue}{b_5}$ and in both drawings, when going along the edge from $\textcolor{red}{r_1}$ to $\textcolor{blue}{b_5}$, it crosses first  $\textcolor{red}{r_2}\textcolor{blue}{b_4}$ and then $\textcolor{red}{r_2}\textcolor{blue}{b_6}$.
}
\label{app:fig:CEP_CR_CO_K2n}
\end{figure}

The drawings in Figure~\ref{app:fig:CEP_CR_CO_K2n} show labeled simple drawings of $K_{\textcolor{red}{2},\textcolor{blue}{6}}$ that are CO-isomorphic (and thus CE-isomorphic) and CR-isomorphic, but the rotation system is different. There cannot be a labeling such that the drawings are RS-isomorphic and CE-isomorphic, for the following reasons.
There is one unique edge that has two crossings ($\textcolor{red}{r_1}\textcolor{blue}{b_5}$). In the left drawing, the rotation around its incident vertex on the smaller bipartition class ($\textcolor{red}{r_1}$) that edge comes between two completely uncrossed edges. In the right drawing it does not.  
Theses drawings can be extended to any $K_{\textcolor{red}{2},\textcolor{blue}{n}}$ with $\textcolor{blue}{n}\geq 3$ by making copies of vertex $\textcolor{blue}{b_4}$ as shown with $\textcolor{blue}{b'}$ and the dashed edges in Figure~\ref{app:fig:CEP_CR_CO_K2n}.

\end{document}